\newif\ifabstract
\abstracttrue
 \abstractfalse 
\newif\iffull
\ifabstract \fullfalse \else \fulltrue \fi

\documentclass[11pt]{article}
\usepackage{amsfonts}
\usepackage{amssymb}
\usepackage{amstext}
\usepackage{amsmath}
\usepackage{xspace}
\usepackage{theorem}
\usepackage{graphicx}
\usepackage{url}
\usepackage{graphics}
\usepackage{colordvi}
\usepackage{colordvi}
\usepackage{subfigure}

\usepackage{natbib}
\usepackage{wasysym}

\usepackage{moreverb} 
\usepackage{textcomp} \usepackage{epsfig}
\usepackage{color} \usepackage{upgreek}
\usepackage{bbm}

\usepackage{subeqnarray}

\advance \oddsidemargin by -0.8in
\newcommand{\myparskip}{3pt}
\parskip \myparskip

\newcommand\Rey{\mbox{\textit{Re}}}  
\newcommand\Pran{\mbox{\textit{Pr}}} 

\newsavebox{\astrutbox}
\sbox{\astrutbox}{\rule[-5pt]{0pt}{20pt}}

\begin{document}

\title{Optimal transient growth in thin-interface internal solitary waves}
\author{Pierre-Yves Passaggia\thanks{Department of Marine Sciences, University of North Carolina, Chapel Hill, NC 27599, USA. Email: passaggia@unc.edu {\tt passaggia@unc.edu}.}
\and Karl R. Helfrich\thanks{Department of Physical Oceanography, Woods Hole Oceanographic Institution, Woods Hole, MA 02543, USA. Email: {\tt khelfrich@whoi.edu}.} 
\and Brian L. White\thanks{Department of Marine Sciences, University of North Carolina, Chapel Hill, NC 27599, USA. Email: {\tt bwhite@unc.edu}.}}

\maketitle

\thispagestyle{empty}

\begin{abstract}
The dynamics of perturbations to large-amplitude Internal Solitary Waves (ISW) in two-layered flows with thin interfaces is analyzed by means of linear optimal transient growth methods. Optimal perturbations are computed through direct-adjoint iterations of the Navier-Stokes equations linearized around inviscid, steady ISWs obtained from the Dubreil-Jacotin-Long (DJL) equation.
Optimal perturbations are found as a function of the ISW phase velocity $c$ (alternatively amplitude) for one representative stratification. These disturbances are found to be localized wave-like packets that originate just upstream of the ISW self-induced zone (for large enough $c$) of potentially unstable Richardson number, $Ri < 0.25$. They propagate through the base wave as coherent packets whose total energy gain increases rapidly with $c$. The optimal disturbances are also shown to be relevant to DJL solitary waves that have been modified by viscosity 
representative of laboratory experiments. The optimal disturbances are compared to the local WKB approximation for spatially growing Kelvin-Helmholtz (K-H) waves through the $Ri < 0.25$ zone. The WKB approach is able to capture properties (e.g., carrier frequency, wavenumber and energy gain) of the optimal disturbances except for an initial phase of non-normal growth due to the Orr mechanism. The non-normal growth can be a substantial portion of the total gain, especially for ISWs that are weakly unstable to K-H waves. The linear evolution of Gaussian packets of linear free waves with the same carrier frequency as the optimal disturbances is shown to result in less energy gain than found for either the optimal perturbations or the WKB approximation due to non-normal effects that cause absorption of disturbance energy 
into the leading face of the wave. Two-dimensional numerical calculations of the nonlinear evolution of optimal disturbance packets leads to the generation of large-amplitude K-H billows that can emerge on the leading face of the wave and that break down into turbulence in the lee of the wave. The nonlinear calculations are used to derive a slowly varying model of ISW decay due to repeated encounters with optimal or free wave packets.
\end{abstract}


\section{Introduction}
The transfer of energy from large to small scales in density stratified fluids such as the oceans, lakes and the atmosphere is known to be driven in large part by internal gravity waves \citep{StaquetS:02}. One particularly energetic type of internal wave found in each of these systems is the Internal Solitary Wave (ISW); see the recent reviews by \citet{HelfrichM:06} and \citet{Grimshaw:10}. In the ocean these waves can be quite large, with wave amplitudes, $\eta_{MAX}$, in excess of 
$240$ meters in some cases \citep{Huang16} and, equally notable, nonlinearity $\alpha = \eta_{MAX}/h \approx 5$, 
where $h$ is an appropriate depth scale \citep{Stanton:98}.
There are several mechanisms that generate oceanic ISWs with the most common being the interaction of barotropic tide with a localized topographic feature. This leads to the radiation of an internal tide that subsequently steepens through nonlinearity to produce one or more ISWs. See \citet{Jackson:12} for a recent review on this and other ISW generation mechanisms. Once produced, these large-amplitude ISWs can propagate for very long distances. A striking example from the South China Sea where the waves emerge from the westward propagating internal tide generated in the Luzon Strait \citep{Alford:15}. The waves have amplitudes $> 100$ meters and travel hundreds of kilometers across the South China Sea to the continental shelf where they shoal, break and dissipate \citep{StLaurent:11}. The resulting vertical turbulent mixing from wave shoaling can be significant \citep{Sandstrom:84,Shroyer:10}.

The waves are also subject to internal instabilities as they propagate in constant depth \citep{Moum:03, Shroyer:10}. 
\citet{ZhangA:15} analyzed 6 months of in-situ data from the Washington continental shelf that produced records of nearly $600$ individual ISWs, with over $120$ exhibiting instabilities and turbulent dissipation. They categorized the waves based on a Froude number, $Fr=u_s/c$, where $u_s$ is the near-surface fluid velocity and $c$ is the wave phase speed, and identified two types of ISWs, each associated with a different type of instability. The first wave type had $Fr<1$ and was characterized by a thin region of Richardson number, 
$$Ri = \frac{N^2}{(\partial u/\partial z)^2}~,$$ 
less than $0.25$, a necessary but not sufficient condition for Kelvin-Helmholtz (K-H) instability. Here $N$ is the local Brunt-V\"ais\"al\"a, or buoyancy, frequency and $\partial u/\partial z$ is the local wave-induced vertical shear. These waves were subject to an instability that produced large amplitude K-H billows \citep[see also][]{Moum:03} and dissipation localized within and downstream of the interfacial region of low $Ri$. The second type of wave had $Fr > 1$ and was associated with the formation of a trapped vortex core \citep[c.f.][]{Davis:67,HelfrichW:10} and was  characterized by turbulent mixing within the overturning core. The interfacial K-H instability was significantly more common and is the focus of the work presented here. Despite the extensive data set, they could not find a clear relationship between the occurrence of instabilities and parameters such as wave steepness, stratification, or mean flow, except that unstable waves tend to be more energetic (i.e., have larger amplitudes and phase speeds). From observations on Scotian Shelf, \citet{Sandstrom:84} did deduce that a minimum Richardson number $Ri_{min} \lesssim 0.1$ was necessary for shear instabilities. Laboratory experiments \citep{Carr:08, Fructus:09,Luzzatto-FegizH14} and numerical simulations \citep{BaradF:10,Almgren:12} have also indicated that $Ri_{min} \lesssim 0.1$ is required for observable K-H billows.

However, a condition for instability based solely on $Ri_{min}$ does not appear to be sufficient. From their laboratory experiments in nearly two-layered stratification with a thin interfacial region, \citet{Fructus:09} suggested an alternative criterion for ISWs to be unstable, namely that  $L_{Ri}/\xi \approx 0.86$, where $L_{Ri}$ is the half-length of the $Ri < 0.25$ zone and $\xi$ is the half-length of the ISW (measured between the wave crest and where the interfacial displacement equals $\eta_{MAX} /2$. However, as shown below, $Ri_{min}$, $L_{Ri}$, and $\xi$ are functions of $c$ for a given background stratification, and so are not independent. This implicated the length of the potentially unstable zone, that is, the spatial or temporal extent available for growing K-H instability, as a critical consideration. Subsequent attempts to define an instability criterion based on the length or the time spent by a perturbation in the region with $Ri < 0.25$, using linear growth rates predicted by the Taylor-Goldstein equation from either either a temporal analysis \citep{TroyK05,Fructus:09,BaradF:10,Almgren:12} or a spatial analysis \citep{LambF11, CamassaV12} have yielded similar results. For example, \citet{TroyK05} found that 
$Ri_{min} \lesssim 0.1$ and $\bar{\omega}_i T_W > 5$. Here $\bar{\omega}_i$ is the average temporal growthrate in the $Ri < 0.25$ region and  $T_W$ ($\approx 2L_{Ri}/c)$ is a measure of the time it takes a disturbance to propagate through the zone. Similarly, \citet{LambF11} found $Ri_{min} \lesssim 0.1$, $L_{Ri}/\xi > 0.8$, and $2 \bar{k}_i L_{Ri} > 4$ are necessary. Here $\bar{k}_i$ is the average spatial growthrate in the low $Ri$ region. In all these cases the criteria rest on the appearance of finite-amplitude K-H billows. Not surprisingly then, the criteria are sensitive to the properties (e.g. frequency, amplitude, etc.) of the perturbations \citep{Almgren:12,LambF11}.

These criteria were recently questioned by \citet{CamassaV12} where the response of large-amplitude ISWs in nearly two-layered stratifications to infinitesimal disturbances was shown to be critically connected with the variable and non-parallel structure of the $Ri < 0.25$ region. They demonstrated that this could promote the absorption of perturbation energy into the spatially-varying vertical shear field as a disturbance entered the wave and also to the clustering of local eigenvalues along the wave. Their study promotes an energetic coupling among neutral modes stronger than what may be expected to occur in 
parallel or slowly varying flows and gives rise to multi-modal transient dynamics of the kind often referred to as non-normality effects. Additionally, the ISW flow configuration is characteristic of globally stable but strongly non-normal system, also known as a noise amplifier \citep{Chomaz,SchmidH01}, in which perturbations localized in the pycnocline grow exponentially as they travel along the ISW and leave the ISW without inducing self-sustained instabilities \citep{LambF11,CamassaV12}.  

Motivated by these considerations, the transient growth of linear instabilities in ISWs is investigated by computing optimal initial disturbances that maximize the energy of the perturbation for a given time horizon \citep{Schmid:2007,LuchiniB:2014} where the baseflow is an ISW found by solution of the Dubreil-Jacotin-Long (DJL) equation. This approach makes no assumptions regarding the perturbation properties and does not require parallel or even a slowly-varying background state. It provides an upper bound on the energy growth generated by a perturbation with an infinitesimal amplitude. The results from the optimization procedure are then compared with a local WKB approach to transient growth, based on the assumption of a weakly non-parallel base flow, together with initial short-time transient growth mechanisms and the growth experienced by packets of free linear internal waves. 

The questions that we seek to address are: What are the necessary and sufficient conditions for a wave to be unstable? What perturbations grow, and under what conditions do K-H billows form? How does this depends on the background wave field and other perturbations? How much energy is extracted from the ISW? For parallel shear flows, the fastest growing modes are two dimensional. Is this the case in ISWs with their curved and spatially varying shear layers? And how do these results impact the issue of dissipation, or decay, of ISWs?

The following manuscript is organized as follows. The governing equations, the DJL model, and solitary wave solutions are introduced in \S\ref{sec:geom}. The transient growth optimization problem is derived and explored in \S \ref{sec:optim}. This is compared to an analysis of slowly varying (WKB) linear spatial K-H instability in \S \ref{sec:WKBJ}, where the effects of non-normal growth are highlighted. The linear evolution of packets of free waves is examined in \S \ref{sec:free_waves}. The nonlinear development of both types of disturbances and the primary ISW are considered in \S\ref{sec:nonlin}. Conclusions are drawn in \S \ref{sec:conclusion}.

\section{DJL solitary waves}\label{sec:geom}

\subsection{Equations of motion}\label{subsec:flow-config}

The fluid motion is governed by the incompressible Navier-Stokes system in the
Boussinesq approximation, written formally, in non-dimensional form as
\begin{equation}
\label{NS}
\begin{split}
& ~~~~~~~~~\mathbf{E}\partial_t\mathbf{q} = \mathbf{F}(\mathbf{q}, \Rey, Sc )~, \\
& \mathbf{F}(\mathbf{q}, \Rey, Sc ) = \left[- (\mathbf{u}\cdot\nabla)\mathbf{u} -\nabla p - s\mathbf{e_z}  + \frac{1}{\Rey}\nabla^2\mathbf{u};\,
\nabla\cdot\mathbf{u};\, - (\mathbf{u}\cdot\nabla)s + \frac{1}{Sc\Rey}\nabla^2 s\right] \\
\end{split}
\end{equation}
for the velocity $\mathbf{u}(x,z,t)$, the density $s(x,z,t)$ and the pressure field $p(x,z,t)$. The solution vector is $\mathbf{q}=(\mathbf{u},p,s)^T$ and $\mathbf{E}$ is the projection operator onto the velocity and density fields such that $\mathbf{E}\mathbf{q}=(u,v,w,0,s)^T$. 
The variables have been scaled using the reduced gravity $g^\prime=g(\rho_{b}-\rho_{0})/\rho_0$, where $g$ is the acceleration of gravity, 
$\rho_{b}$ and $\rho_{0}$ are, respectively, the densities at the bottom and top of the domain of depth $H$. 

The non-dimensional quantities are given by
\begin{equation}
(x,z) = \frac{(x^*,z^*)}{H},\quad t = t^*\sqrt{\frac{H}{g'}},\quad(\mathbf{u},c) = \frac{(\mathbf{u}^*,c^*)}{\sqrt{g'H}},\mbox{ and }~ s = \frac{\rho^*-\rho_0}{\Delta \rho},
\label{nondim} 
\end{equation}
where the $*$ denotes a dimensional quantity. Note that $\rho^*$ is the density and $s$ is a scaled departure from $\rho_0$.
The Reynolds number $\Rey = \sqrt{g'H^3}/\nu$ and  $Sc = \nu/\kappa$ is the Schmidt number. Here  $\nu$ is the kinematic 
viscosity of the fluid and $\kappa$ is the diffusivity of the stratifying agent.

The two-dimensional flow domain is $-L \leq x \leq L$, $0 \leq z \leq 1$.
Adiabatic and free slip conditions, $\left[\partial u/\partial  z, w, \partial s/\partial  z\right] = 0$, 
 are imposed on $z=0$ and $1$ and the lateral boundary condition at $x= \pm L$ are taken as a specified upstream inflow and, for the numerical solutions, a downstream outflow with an advective-type condition 
\begin{equation}
\frac{\partial\mathbf{u}}{\partial t}=c\frac{\partial\mathbf{u}}{\partial x}, \quad \frac{\partial s}{\partial t}=c\frac{\partial s}{\partial x},
\end{equation}
is employed.

\subsection{DJL theory for internal solitary waves}

The finite-amplitude internal solitary waves that form the base state for the stability analysis are obtained from the Dubreil-Jacotin-Long (DJL) equation \citep{Dubriel:34, Long:53}. It is an exact reduction of the steady, two-dimensional Euler equations ((\ref{NS}) with $\nu=\kappa=0$). For a Boussinesq fluid of depth $H$, the DJL equation is, in dimensional variables, \citep[c.f.][]{StastnaL:02}
\begin{equation}
\nabla^2\eta + \frac{\bar{N}^2(z-\eta)}{c^2}\eta = 0,
\label{DJL}
\end{equation}
with boundary conditions 
\begin{equation}
\eta(x,0) = \eta(x,H)=0, \quad \eta(\pm\infty,z) \rightarrow 0.
\label{DJL_BC}
\end{equation}
Here $c$ is the wave phase speed, $\eta(x,z)$ is the displacement of an isopycnal from its upstream resting position. The buoyancy frequency $\bar{N}$ of the resting density profile $\bar{\rho}(z)$ is given by
$$
\bar{N}^2(z) = -\frac{g}{\rho_0}\frac{d\bar{\rho}}{dz}, ~~~~~~\bar{\rho}(z) = \rho_0 + \Delta \rho \bar{S}(z).
$$
The scaled background density profile $\bar{S}(z)=[0,1]$. In the non-dimensionalization of (\ref{nondim}) with $\eta$  scaled by $H$, (\ref{DJL}) and (\ref{DJL_BC}) are unchanged except that $\bar{N}^2=-d\bar{S}/dz$. In the frame moving with the wave the streamfunction $\Psi = c(\eta-z)$, the velocities $(U,W) = (\Psi_z,-\Psi_x)$, and the density field is $S(x,z)=\bar{S}(z-\eta(x,z))$. Upper case symbols are used for the ISW fields to distinguish them from the perturbation variables introduced in the next section.

Given $\bar{S}(z)$, a family DJL solutions that branch from the linear long wave with phase speed $c_0$ are obtained for increasing values of $c$ $(> c_0)$ using Newton-Raphson iterations, 
where $c$ is added as a parameter in a pseudo-arclength continuation method \citep{Luzzatto-FegizH14}. Standard second-order finite differences are used for the Laplacian operator. Iterations are continued until the $L_2$ norm of the corrections is less than $10^{-10}$. Since solitary waves (see below) are symmetric about the wave crest, we take $\partial \eta /\partial x = 0$ at $x=0$ and reduce the domain to $0<x<L$ and take $\eta(L,z) = 0$. $L \ge 6$ is made large enough that the ISW solutions are not affected by the finite size. The calculation is started from a weakly nonlinear solitary wave solution to the Korteweg-de Vries (KdV) equation \citep{HelfrichM:06}. Typically  $250$ cells in $z$ and $500$ in $x$ are used. However, when a particular wave is needed for a calculation, the solution is interpolated onto the desired fine grid and adjusted to convergence by additional Newton-Raphson iterations.  

\begin{figure}
\begin{center}
\includegraphics[width=85mm]{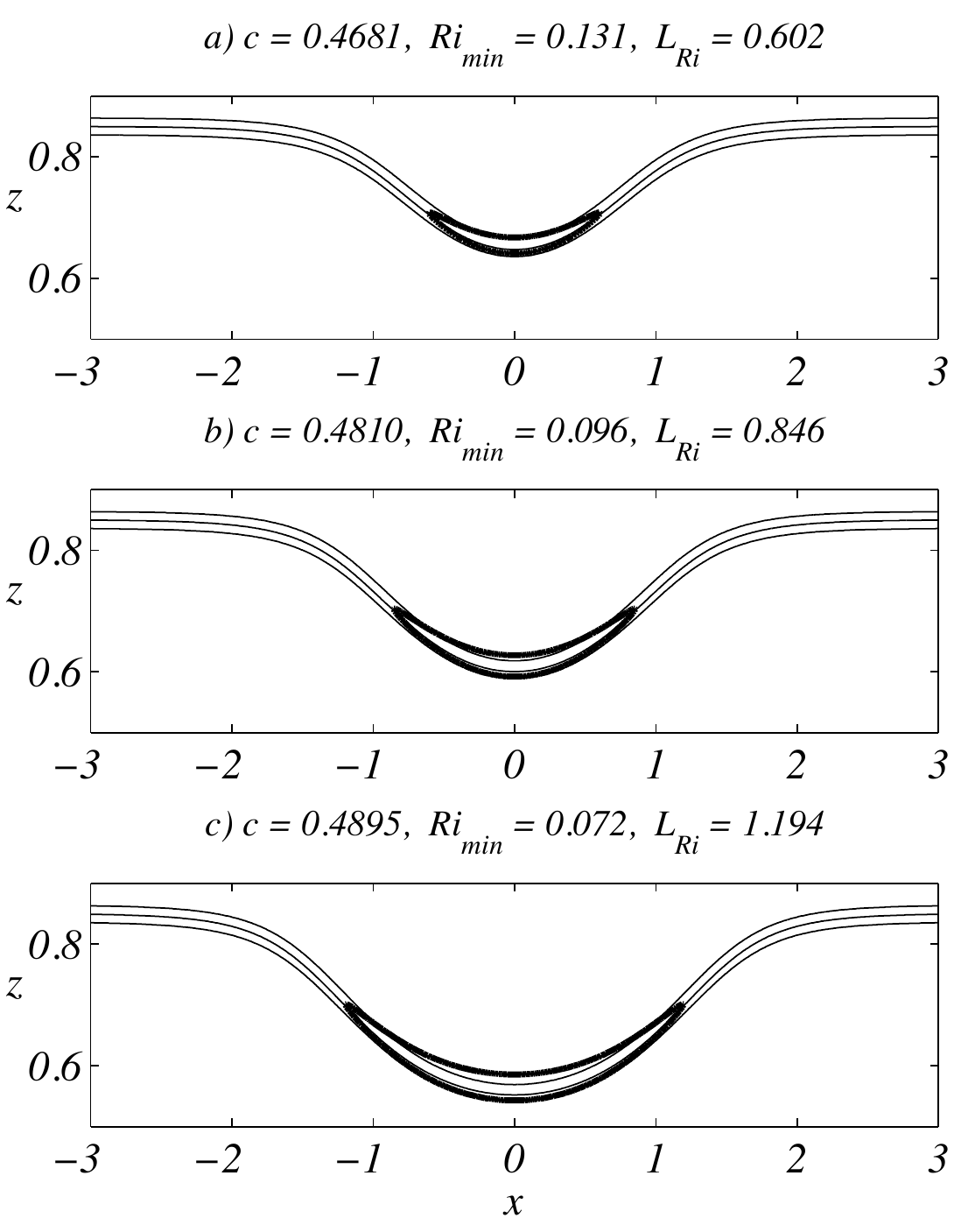}
\caption{Three DJL internal solitary waves for the density profile (\ref{Sbar}) with $z_0 = 0.85$ and $\lambda = 80$. Values for $c$, $Ri_{min}$ and $L_{Ri}$ are indicated. The thin lines are the $S = [0.1, 0.5, 0.9]$ isolines and the heavy line shows the $Ri = 0.25$ contour. Only a portion of the full domain is shown.}
\label{fig_threewaves}
\end{center}
\end{figure}

\begin{figure}
\begin{center}
\begin{subfigure}{}
\includegraphics[width=65mm]{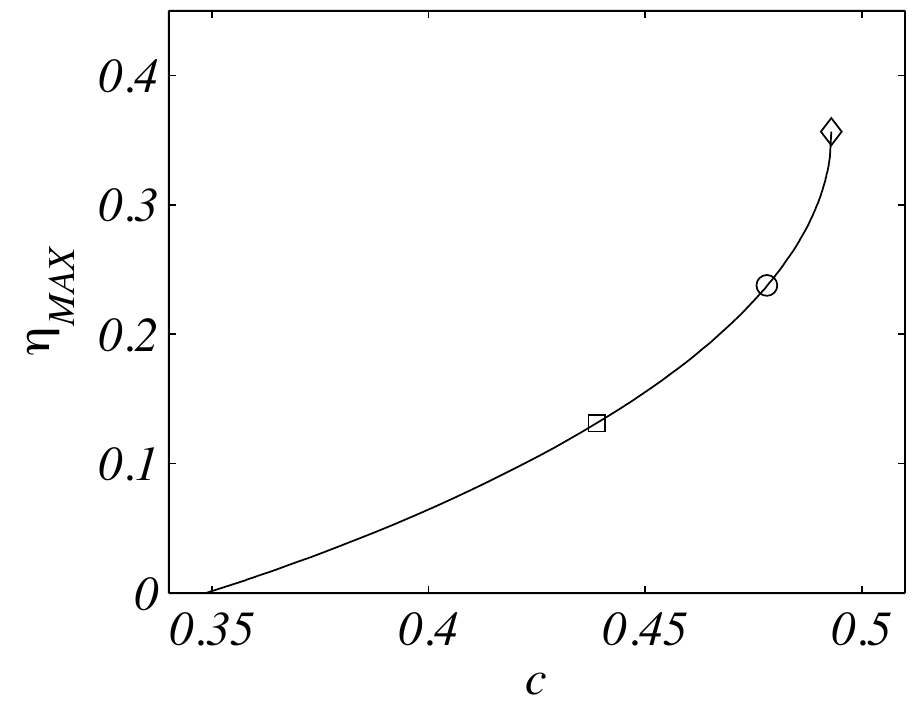}
\put(-130,120){(a)}
\end{subfigure}
\begin{subfigure}{}
\includegraphics[width=65mm]{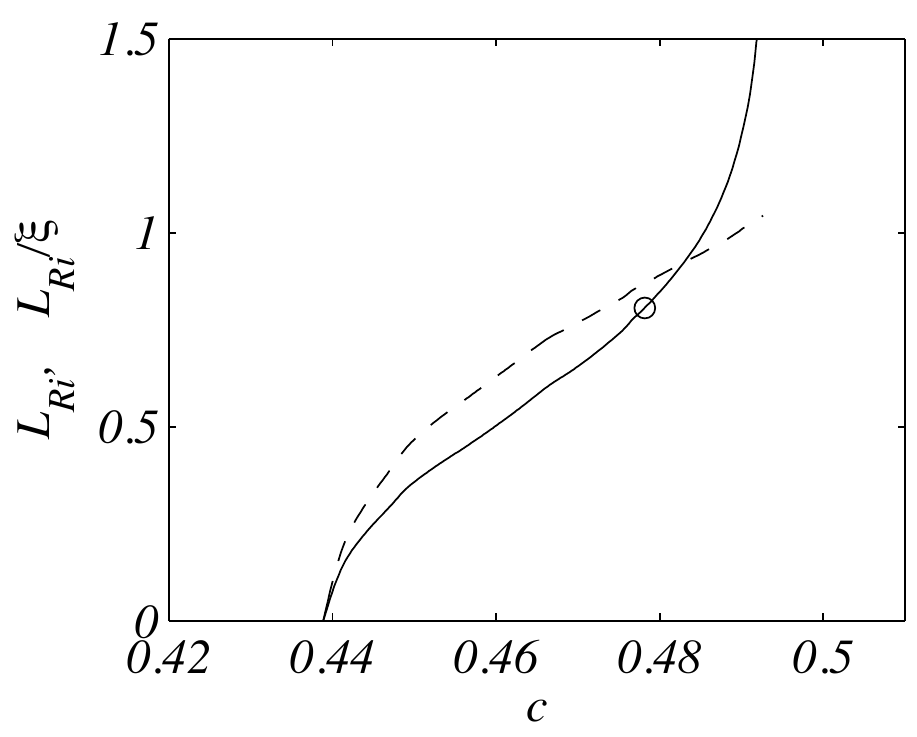}
\put(-130,120){(b)}
\end{subfigure}
\caption{a) $\eta_{MAX}$  and b) $L_{Ri}$ (solid) and $L_{Ri}/\xi$ (dashed) versus $c$ for the DJL solitary wave solutions with $z_0 = 0.85$ and $\lambda = 80$. The square (circle) indicates $Ri_{min} = 0.25$ ($0.1$), the diamond indicates the maximum conjugate state wave.}
\label{etamax_Li_c}
\end{center}
\end{figure}

\subsection{DJL solitary waves}

A two-layer stratification is given by
\begin{equation}
\bar{S}(z) = \frac 1 2 \left(1-\tanh\left[\lambda(z-z_0)\right]\right),
\label{Sbar}
\end{equation}
where $z_0$ is the location of the interface with thickness scale $\lambda^{-1}$. In what follows $z_0 = 0.85$ and $\lambda=80$ are used to produce a thin pycnocline close to the upper boundary. Note that because the interface is in the upper half of the domain, the solitary waves are waves of depression with negative wave amplitudes. However, the amplitude $\eta_{MAX} = |{\rm min}[\eta(0,z)]|$ is defined as the magnitude for convenience. This definition is close to, but not precisely, the maximum displacement of the $S = 0.5$ isopycnal.

Three example DJL solutions are shown in Figure \ref{fig_threewaves} for increasing $c$. The figures show the isopycnals $S(x,z) = [0.1, 0.5, 0.9]$ and the boundary of the $Ri < 0.25$ region. Each of these waves has $Ri_{min} < 0.25$, which always occurs at the wave crest, and decreases as $c$ increases.  The half-length of this region, $L_{Ri}$, amplitude $\eta_{MAX}$, and the wave width, $\xi$, all increase with $c$ (the latter only for this range of $c$). 

The relationship between $c$ and $\eta_{MAX}$ for the full family of ISWs for this stratification is shown in Figure \ref{etamax_Li_c}a. The solutions branch from infinitesimal linear long waves at $c_0=0.3485$ and end at the conjugate state wave \citep[c.f.][]{LambW:98} with $c_{cs}=0.4930$ and $\eta_{MAX}=0.3565$. The flat-crested, infinitely broad conjugate state is found from a one dimensional version of the DJL model following \citet{LambW:98}. In Figure \ref{etamax_Li_c}b shows the behavior of $L_{Ri}$ and $L_{Ri}/\xi$ with $c$. Waves with $Ri_{min} < 0.25$ are found for $c > 0.4389$ and $Ri_{min}=0.1$ at $c = 0.4792$ where $L_{Ri} = 0.807$ and $L_{Ri}/\xi = 0.928$. 

In the following, we consider the dynamics of infinitesimal perturbations, governed by (\ref{NS}), to ISWs from the family of DJL solutions shown in Figures \ref{fig_threewaves} and \ref{etamax_Li_c}. These waves are representative of internal solitary waves on similar thin-interface background stratifications. The aim here is to compare the most amplified transient dynamics 
of two-dimensional wave packets and compare their evolution with the amplification provided by a local asymptotic expansion in terms of unstable normal modes where a separation of scales between the DJL wave and the wave packet is assumed following \citet{CamassaV12}.

\subsection{Viscous adjusted steady states}\label{sec:viscsteadystate}

The question naturally arises whether viscosity modifies the DJL waves sufficiently, at least for laboratory scales, such that the  transient growth results are significantly affected. 
However, viscosity does not allow for an equation of the form of (\ref{DJL})
and viscous DJL waves will be unsteady, although the temporal changes will be slow for large $\Rey$. See, for example, \citet{Grimshaw:03} for an adiabatic approximation for decaying KdV solitary waves. Quasi-steady, viscously adjusted DJL waves can be found using the Selective Frequency Damping (SFD) method \citep{Akervik:06} where the Navier-Stokes equations (\ref{NS}) are coupled to a low-pass temporal filtered solution $\bar{\mathbf{q}}$ (a sliding average of the state solution $\mathbf{q}$) of the form
\begin{equation}
\begin{array}{c}
\partial_t \mathbf{q} = \mathbf{F}(\mathbf{q},Re) - \zeta(\mathbf{q}-\bar{\mathbf{q}}) \\ 
\partial_t \bar{\mathbf{q}} = (\mathbf{q}-\bar{\mathbf{q}})/\chi,
\end{array}
\label{SFD}
\end{equation}
where $\chi^{-1}$ is the width of the temporal filter, and $\zeta$ is the amplitude of the damping applied to the Navier-Stokes equations.
When initialized with a DJL wave, this approach rapidly relaxes the inviscid wave to a viscously adjusted wave as (\ref{SFD}) is integrated in time. The optimal cutoff frequency $\chi=12.7423$ and damping factor $\zeta=0.0534$ 
at $\Rey=10^5$ were computed using the approach of \citep{Cunha:15}. 

\section{The transient growth optimization problem}\label{sec:optim}

\subsection{Direct and adjoint problem formulation}\label{sec:optim1}

The solution is now decomposed between a base flow, the solitary wave, and a perturbation such that
\begin{equation}
\mathbf{q}(\mathbf{x},t) =  \mathbf{Q}(\mathbf{x}) + \mathbf{q}'(\mathbf{x},t),
\end{equation}
where $\mathbf{Q}=(\mathbf{U}, P, S)^T$ is the DJL solution or viscous adjusted DJL wave and $\mathbf{q}' = (\mathbf{u}', p', s')^T$ contains the perturbation velocity field $\mathbf{u}'$, the perturbation pressure $p'$, and the density $s'$. 

The evolution of infinitesimal amplitude perturbations are solution of the linearized Navier-Stokes system, where the second order nonlinear terms ($\mathbf{u}'\cdot\nabla\mathbf{u}', \mathbf{u}'\cdot\nabla s'$) are negligible, becomes
\begin{equation}
\begin{split}
\mathbf{f}(\mathbf{Q}, \mathbf{q}', \Rey, Sc) \equiv  \bigg{[}
& \partial_t \mathbf{u}' + \Phi(x)\Big{(}(\mathbf{U}\cdot\nabla)\mathbf{u}' + (\nabla\mathbf{U})\mathbf{u}'\Big{)} +\nabla p' + s'\mathbf{e_z}  - \frac{1}{\Rey}\nabla^2\mathbf{u}',\\
& \nabla\cdot\mathbf{u}',\\ 
& \partial_t s' + \Phi(x)\Big{(}(\mathbf{U}\cdot\nabla)s' +(\nabla S)\mathbf{u}'\Big{)} - \frac{1}{Sc\Rey}\nabla^2 s' \bigg{]}=0.
\end{split}
\label{direct}
\end{equation}
The sponge layer, 
\begin{equation}
\Phi(x)=\frac{1}{2}[1+\tanh(10\times(x \pm x_0))],
\label{Phi}
\end{equation}
is added to the advection and production terms and is minimum near the inlet and the outlet. It allows the direct solutions to vanish as they approach the lateral boundaries of the domain far from the shear-induced regions of the ISW and without reflection. 
The length of the sponge layer, $x_0=0.73L$, was chosen such that the maximum distance a perturbation can travel is at least twice longer than the unstable region ($Ri<1/4$) of the larger amplitude DJL wave.

In the following optimization problem, we 
seek to maximize an objective function $G(\mathbf{q}'(\mathbf{x},T))$ which is a measure of the energy of the perturbation $\mathbf{q}'$ at a finite time $T$, normalized by the initial energy $E_0$ at time $t=0$ and is given by
\begin{equation}
G(\mathbf{q}'(T)) = \frac{E(T)}{E_0} =  \left[\frac 1 2 \int_{\Omega}  \left( \mathbf{u}'(T)\cdot\mathbf{u}'(T) + \frac{(s'(T))^2}{N^2_*}\right)\,\mbox{d}\mathbf{x}\right]/E_0,
\label{energy_dal}
\end{equation}
where $\Omega$ denotes the computational domain.
In the present linear analysis  $E_0=1$. Here $N_*^2$ is the maximum Brunt-V\"ais\"al\"a frequency of the baseflow $S$, chosen to avoid division by zero in regions where  $N^2(\mathbf{x})$ approaches zero.

This optimization problem is constrained by the linearized Navier-Stokes equations (\ref{direct}), which can be solved by introducing Lagrange multipliers $\mathbf{q}^+$ for the solution vector, $\mathbf{q}_0^+$ for the initial condition and $E_0^+$ for the initial energy.
Hence the Lagrangian is given by
\begin{equation}
\begin{split}
\mathcal{L}(\mathbf{q}',\mathbf{q}^+) &= G -<<\mathbf{f}(\mathbf{U},S,\mathbf{q}',\Rey\,Sc),\mathbf{q}^+>> - <\mathbf{g}(\mathbf{u}',s', 
\mathbf{u}'_0, s'_0),(\mathbf{u}_0^{+},s_0^{+})> \\ &- h(\mathbf{u}'_0, s'_0, E_0)E_0^+,
\end{split}
\label{Lag}
\end{equation}
and is to be rendered stationary. The scalar product $<\mathbf{a}\cdot\mathbf{b}>$ is defined by the spatial integral $\int_{\Omega} \mathbf{a}\cdot\mathbf{b}\,\mbox{d}\mathbf{x}$ whereas  $<<\mathbf{a}\cdot\mathbf{b}>>$ is defined by the double integral $\int_{0}^{T}\int_{\Omega} \mathbf{a}\cdot\mathbf{b}\,\mbox{d}\mathbf{x}\mbox{d}t$  where the optimization window is taken in the time interval $[0,T]$. The constraints for the initial state $\mathbf{g}(\mathbf{u}',s', \mathbf{u}'_0, s'_0)$ and the initial amplitude $h(\mathbf{u}'_0, s'_0, E_0)$ associated with $\mathbf{u}'_0$ and $E_0$ respectively are
\begin{equation}
\begin{split}
&\mathbf{g}(\mathbf{u}',s', \mathbf{u}'_0, s'_0) = (\mathbf{u}'_0-\mathbf{u}'(0) , s'_0-s'(0)) = 0,\\
&h(\mathbf{u}'_0, s'_0, E_0) = \frac{1}{2}\int_{\Omega} \mathbf{u}'_0\cdot\mathbf{u}'_0 + s_0^{'2}/N^2_*\mbox{d}\mathbf{x}-E_0=0.
\end{split}
\end{equation}

Taking variations of the Lagrangian (\ref{Lag}) with respect to the state variable $\mathbf{q}$ and setting the result equal to zero, the adjoint system is equivalent to the one derived by \citet{KaminskiCT14} and is
\begin{equation}
\begin{split}
\mathbf{f}^+(\mathbf{Q}, \mathbf{q}^+, \Rey, Sc) \equiv \bigg{[}
& -\partial_t \mathbf{u}^+ - \Phi(x)\Big{(}(\mathbf{U}\cdot\nabla)\mathbf{u}^+ + (\nabla\mathbf{U})^T\mathbf{u}^+ + (\nabla S)s^+\Big{)} +\nabla p^+ - \frac{1}{\Rey}\nabla^2\mathbf{u}^+,\\
& -\nabla\cdot\mathbf{u}^+,\\ 
& -\partial_t s^+ - \Phi(x)\Big{(}(\mathbf{U}\cdot\nabla)s^+ - w^+\Big{)} - \frac{1}{Sc\Rey}\nabla^2 s^+ \bigg{]} = 0.
\end{split}
\label{adjoint}
\end{equation}
The boundary conditions of the direct-adjoint equations are determined by the boundary terms $B$ remaining from the integrations by parts leading to (\ref{adjoint}) and are
\begin{equation}
\begin{split}
B &= \int_{0}^{T}\int_{\partial\Omega} \bigg{[}
  \frac 1 \Rey (\nabla{\mathbf{u}}\cdot\mathbf{n})\mathbf{u}^+ 
- \frac 1 \Rey (\nabla\mathbf{u}^+\cdot\mathbf{n}){\mathbf{u}}
-\Phi(x)(\mathbf{U\cdot\mathbf{n}})\mathbf{u}^+\cdot{\mathbf{u}} 
+ (p^+\mathbf{n})\cdot{\mathbf{u}}\\
&- \mathbf{u}^+\cdot({p}\mathbf{n})
+ \frac{1}{\Rey Sc}s^+(\nabla{s}\cdot\mathbf{n})
- \frac{1}{\Rey Sc}(\nabla s^+\cdot\mathbf{n}){s}
- \Phi(x)(\mathbf{U}\cdot\mathbf{n})s^+\cdot{s}
\bigg{]} \mbox{d}A \mbox{d}t\\
&+ \int_{\Omega} \big{[}\mathbf{u}^+\cdot{\mathbf{u}}\big{]}_{0}^{T} + \big{[}s^+\cdot{s}\big{]}_{0}^{T}\mbox{d}\Omega \mbox{d}\mathbf{x},
\end{split}
\label{BT}
\end{equation}
where $A$ denotes the boundary part of the computational domain $\Omega$.  
These boundary terms have to be canceled in order to insure compatibility between the direct system and its adjoint \citep{LuchiniB:2014}. Choosing the boundary conditions for the perturbation such that
\begin{equation}
\mathbf{u}'\cdot\mathbf{n}\Big{|}_{x=\pm L}=0,\quad s'\Big{|}_{x=\pm L}=0,\quad\frac{\partial\mathbf{u}'}{\partial\mathbf{n}}\Big{|}_{z=0,1}=0,
\quad\frac{\partial s'}{\partial\mathbf{n}}\Big{|}_{z=0,1}=0,
\label{bc_direct}
\end{equation}
leads to vanishing of $B$ (\ref{BT}) provided that the adjoint boundary conditions
\begin{equation}
\mathbf{u}^{+}\cdot\mathbf{n}\Big{|}_{x=\pm L}=0,\quad s^{+}\Big{|}_{x=\pm L}=0,\quad\frac{\partial\mathbf{u}^{+}}{\partial\mathbf{n}}\Big{|}_{z=0,1}=0,
\quad\frac{\partial s^{+}}{\partial\mathbf{n}}\Big{|}_{z=0,1}=0,
\label{bc_adjoint}
\end{equation}
are imposed. The direct flow field $(\mathbf{u}',s')$ is obtained through time marching from $0$ to $T$ and it enters the adjoint system 
at time $T$ with
\begin{equation}
\mathbf{u}^+(T) = \mathbf{u}'(T) \quad\mbox{and}\quad s^+(T) = N^2_* s'(T), \end{equation}
which is to be solved backward in time from $T$ to $0$ in a Direct-Adjoint-Loop (DAL) procedure \citep{Schmid:2007}.

Taking variations of the Lagrangian (\ref{Lag}) with respect to the initial solution $\mathbf{q}_0$ and using the last integral in (\ref{BT}), the expression for the gradient of the objective function is
\begin{equation}
\nabla_{\mathbf{q}_0}G(\mathbf{q}_0) = ( \mathbf{u}^+(0)-E_0^+\mathbf{u}_0, 0, s^+(0)/N^2_* -E_0^+s_0/N^2_* ),
\label{gradient}
\end{equation}
with
\begin{equation}
\quad E_0^+ = \sqrt{\int_\Omega \mathbf{u}^+(0) \cdot \mathbf{u}^+(0) + (s^+(0))^2/N^2_*, \;\mbox{d}\mathbf{x}}
\end{equation}
where the expression for $E_0^+$ is found assuming $\nabla_{\mathbf{q}_0} G = 0$.

\subsection{Numerical methods}\label{sec:num_methods}

Two different numerical methods were used to compute the optimal perturbations. In the case of the viscous adjusted ISWs, 
both the baseflow, through (\ref{SFD}), and the optimal perturbation were computed using the same grid which is a mixed spectral/finite differences type discretization. The streamwise direction 
($x$) 
with $L=3$ is discretized using finite differences with $2001$ discretization points, the vertical direction ($z$) uses 200 Chebyshev polynomials and the pressure is solved using a pressure projection method. The temporal integration is semi-implicit and is performed using a second-order backward Euler type scheme for the diffusion part whereas an explicit second order Adams-Bashforth scheme is used for the advection terms. Further details about the numerical discretization procedure can be found in \citet{marquillie:ehrenstein:numerique} and \citet{marquillie:ehrenstein}.

\begin{figure}
\vskip4mm
\centering
\includegraphics[width=0.6\textwidth]{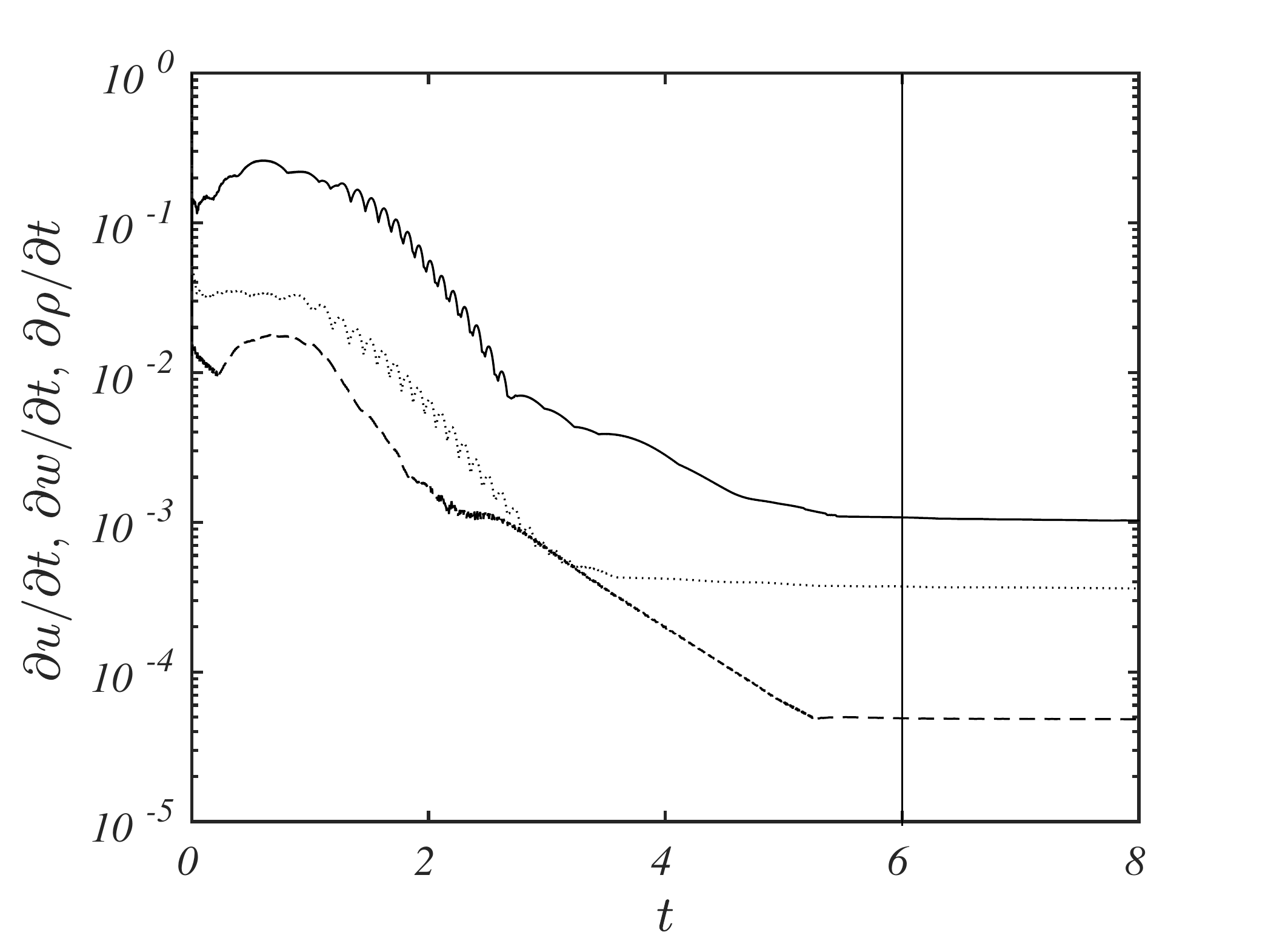}
\caption{Example of the convergence of the infinity norm $||\partial_t u||_\infty$ (- - -), $||\partial_t w||_\infty$ ($\cdots$),$||\partial_t s||_\infty$ (-----) using the SFD algorithm (\ref{SFD}), initialized with the DJL ISW computed at $c=0.4681$ and for the control parameters $\Rey=10^5$ and $\Pran=1$. The vertical line shows when the flow was considered to be at a quasi steady-state.}
\label{conv_SFD}
\end{figure}

Solutions of the SFD system (\ref{SFD}) were considered to be converged to a viscous quasi-steady state when the $L_\infty$-norm of the temporal derivative of each component for the solution vector becomes constant (i.e. where the relaxation imposed by the SFD is only due to viscous effects). Initial transients adjustments are rapidly damped by the SFD and within $6$ to $10$ time units depending on the value of $c$ where the viscous stabilized DJL solutions are found to decay at a rate lower than $10^{-3}$ for the velocity and the density. The convergence of the algorithm is illustrated in Figure \ref{conv_SFD} for a simulation initialized with a DJL wave computed at $c=0.4681$, $\Rey=10^5$ and $Sc=1$. The adjusted wave is then used as a steady base state for an optimal perturbation calculation under the assumption that the timescale for the transient dynamics, $\sim 2L_{Ri}/c$, is much faster than the subsequent changes to the viscously adjusted DJL wave. An adjusted wave develops only slight asymmetry about its crest and was found to propagate at a speed that is only $\approx 4$\% slower than the inviscid wave. Similar to the procedure used to solve the Navier-Stokes system (\ref{NS}) and the stabilized system (\ref{SFD}), the linearized perturbation dynamics and the adjoint system are solved using the same projection method to recover a divergence-free velocity field.

Because of the shorter vertical scale of the interfacial region in an inviscid DJL wave, it proved advantageous to switch to a scheme with a uniform grid of $513$ points in the vertical. Additionally, both the forward and adjoint equations were cast in streamfunction-vorticity form. Spatial derivatives in both $x$ and $z$ were computed with $6th$-order compact finite differences \citep{Lele:92} and the temporal integration uses a $3rd$-order Runga-Kutta method.

We note that for both the inviscid and viscously adjusted DJL waves the domain with total length $2L = 6$ was adequate to eliminate effects of the boundaries because of both the use of the windowing function (\ref{Phi}) and, as shown below, because the optimal disturbances were localized near the $Ri < 0.25$ zone. As a consequence, many of the inviscid DJL wave cases were conducted with no windowing, $\Phi(x)=1$, and periodic conditions in $x$ with no discernible effect. Tests with $L=4$ also showed no influence of the domain length.

Finally, the optimization DAL procedure was considered converged when the $L_2$ norm of the energy difference between two iterates was smaller than $10^{-3}$ and when the $L_2$ norm between two iterates of the gradient (\ref{gradient}) was smaller than $10^{-2}$ which were typically achieved after $12$ to $15$ iterations.

\subsection{Optimal transient growth results}\label{sec:optim4}

\begin{table}
\centering
\begin{tabular}{c c c c c}
\hspace{6mm} $c$ \hspace{6mm} & \hspace{3mm} $\eta_{MAX}$ \hspace{3mm} & \hspace{3.5mm} $Ri_{min}$ \hspace{3.5mm} & \hspace{4mm} $L_{Ri}$  \hspace{4mm} &  \hspace{3mm} $L_{Ri}/\xi$  \hspace{3mm} \\
\hline \hline
0.4442 & 0.142 & 0.230  & 0.188 & 0.250 \\
0.4579 & 0.174 & 0.167  & 0.451 & 0.570 \\
0.4681 & 0.229 & 0.131  & 0.602 & 0.715 \\
0.4810 & 0.249 & 0.096  & 0.846 & 0.876 \\
0.4895 & 0.298 & 0.072  & 1.194 & 0.989 \\
0.4925 & 0.333 & 0.061  & 1.671 & 1.033 \\
 \end{tabular}
\caption{Properties of DJL solitary waves used in the optimal perturbation calculations.} 
\label{tab:cISW}
 \end{table}

\begin{figure}
\begin{center}
\includegraphics[width=80mm]{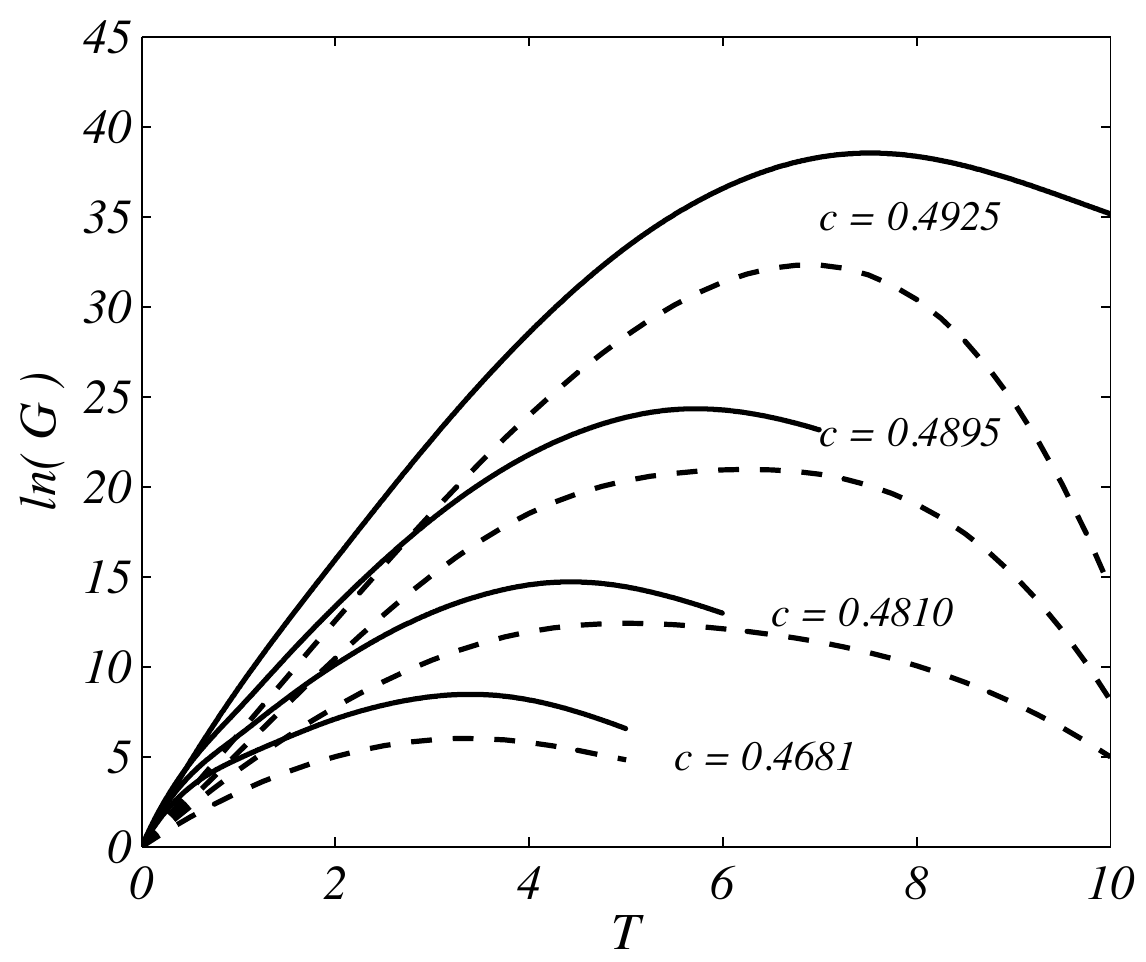}
\caption{Linear transient optimal growth gain $G$ versus the integration time $T$ for $c$ as indicated. 
The gain is found for $\Rey = 10^5$ and $Sc = 1$. The solid (dashed) line is for the inviscid ($Re=10^5$ adjusted) DJL wave base state. }
\label{fig_logG_vs_T}
\end{center}
\end{figure}

Optimal perturbations were computed from the direct-adjoint system  (\ref{direct}-\ref{adjoint}) for six DJL internal solitary waves from the solution family in Figure \ref{etamax_Li_c}. The wave properties $c$, $\eta_{MAX}$, $Ri_{min}$, $L_{Ri}$, and $L_{Ri}/\xi$ are given in Table \ref{tab:cISW}.  
Figure \ref{fig_logG_vs_T} shows the dependence of the natural logarithm of optimal gain
\begin{equation}
\ln[G(T)] = \ln\left[\frac{E(T)}{E(0)}\right]
\label{optG}
\end{equation} 
versus the integration period $T$ for several of these waves. The perturbations were computed for $\Rey=10^5$ and $Sc=1$. This Reynolds number is representative of laboratory experiments in water ($\nu\approx10^{-6}$ m$^2$ s$^{-1}$ , $\Delta\rho/\rho_0=10^{-2}$, and $H\approx 0.5$ m) close to those by \citet{Carr:08,Carr:17}. The choice $Sc=1$ is not correct, but used for numerical convenience. The figure shows $\ln[G(T)]$ for both the inviscid and viscously-adjusted ISW base states at the same $c$. In all cases the gain was found to reach a maximal value, $\ln(G_{MAX})$, at a time $T_{MAX}$. While viscous effects on the primary solitary wave lead to reduced gains and slightly different optimal times, the overall behavior is unchanged.  The maximal linear gain grows dramatically with $c$, with $\ln(G_{MAX}) \approx 40 $ at $c = 0.4925$ for the inviscid DJL wave and $> 30$ for its viscously adjusted counterpart. These correspond to energy gains of up to $10^{17}$, indicating just how unstable these waves can be. It also suggests that the nonlinear evolution of the optimal perturbations needs to be assessed. The optimal time scaled with $T_{MAX} \approx 2L_{Ri}/c$, the time for the perturbation to travel through the potentially unstable zone. This is consistent with \citep{CamassaV12} who argued that thin-interface ISWs are convectively, but not globally, unstable. 

\begin{figure}
\begin{center}
\includegraphics[width=70mm]{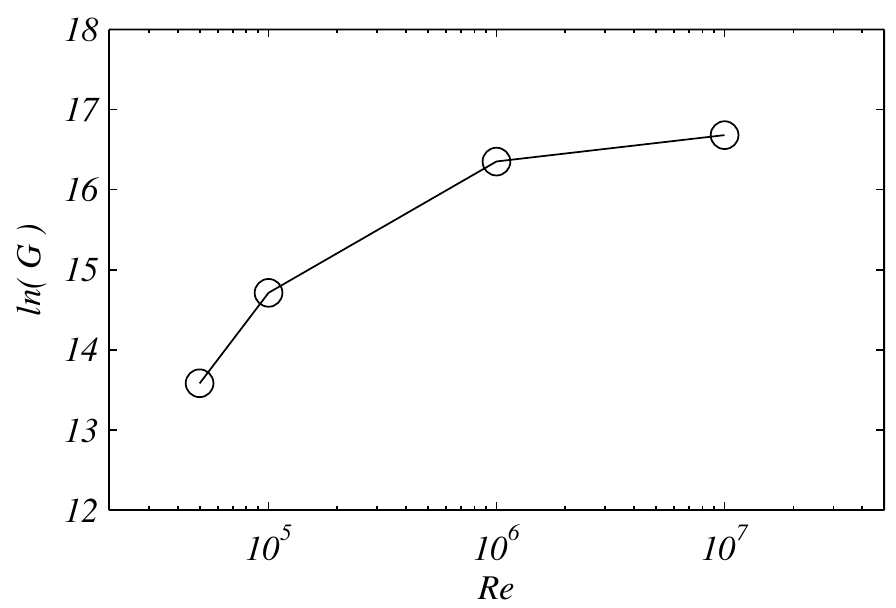}
\caption{$G$ at $T=4.43$ versus $Re$ for $c = 0.4810$ and the inviscid DJL wave.}
\label{fig_logG_vs_Re}
\end{center}
\end{figure}

Figure \ref{fig_logG_vs_Re} shows the effect of $\Rey$ on the perturbation energy gain, $G$ at $T=4.43$ for the $c=0.4810$ inviscid DJL. The gain jumps about one order of magnitude from $\Rey=10^5$ to $10^7$, although the change is small for $\Rey>10^6$, suggesting an inviscid asymptote of $\ln(G) \sim 17$.

\begin{figure}
\begin{center}
\includegraphics[width=70mm]{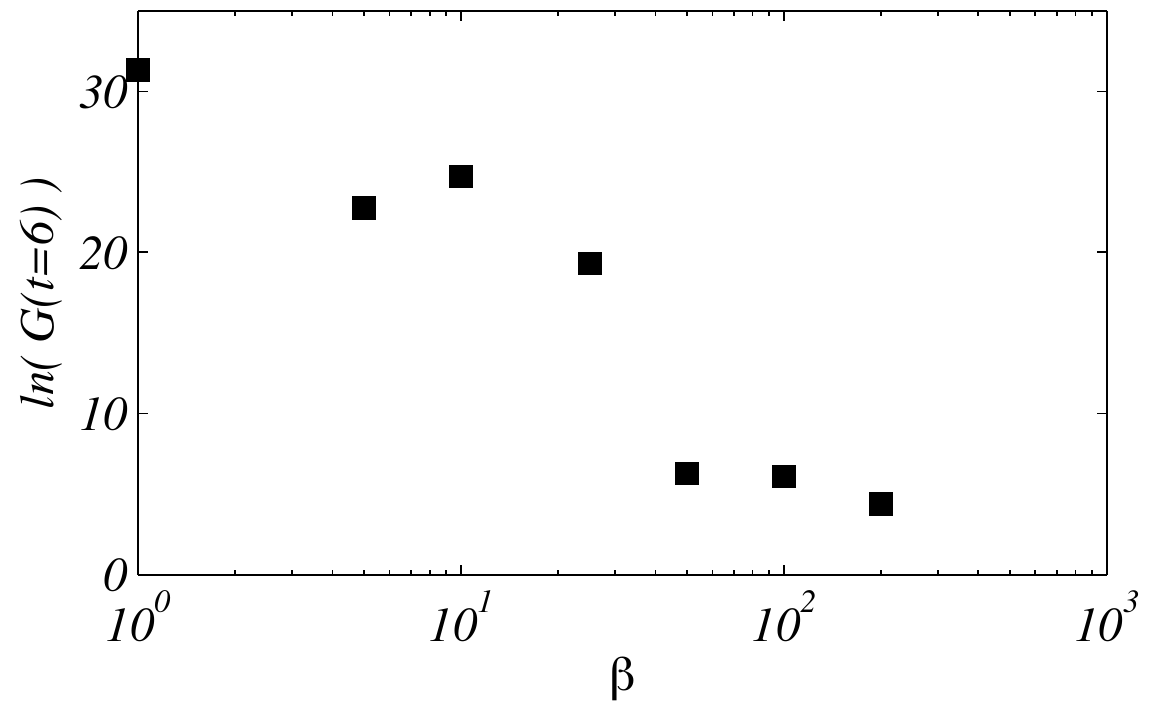}
\caption{$G$ at $T=6$ versus $\beta$, the transverse wavenumber, for the $Re = 10^5$ adjusted DJL wave with $c = 0.4925$.}
\label{fig_logG_vs_beta}
\end{center}
\end{figure}

In shear flows, two-dimensional perturbation are expected to produce the largest gain as a consequence of Squire's theorem \citep{Squire:33,SchmidH01}. This is confirmed here by employing a Fourier decomposition in the transverse ($y$) direction to 
add a three-dimensional perturbation of the form $\mathbf{q}'=\tilde{\mathbf{q}}(x,z)e^{i\beta y}$, where $\beta$ 
is the transverse wavenumber. Linear gains were computed for $T=6$ with the  $c = 0.4925$, $\Rey = 10^5$ adjusted DJL wave (see Figure \ref{fig_logG_vs_beta}). As expected, the optimal gain $G$ decreases with increasing $\beta$. However, it is worth noting that a large gain, $\ln(G)\approx 20$, can still be achieved for $\beta\approx20$. As shown below, this transverse scale is comparable to the longitudinal scale of the optimal disturbance.

\begin{figure}
\begin{center}
\includegraphics[width=60mm]{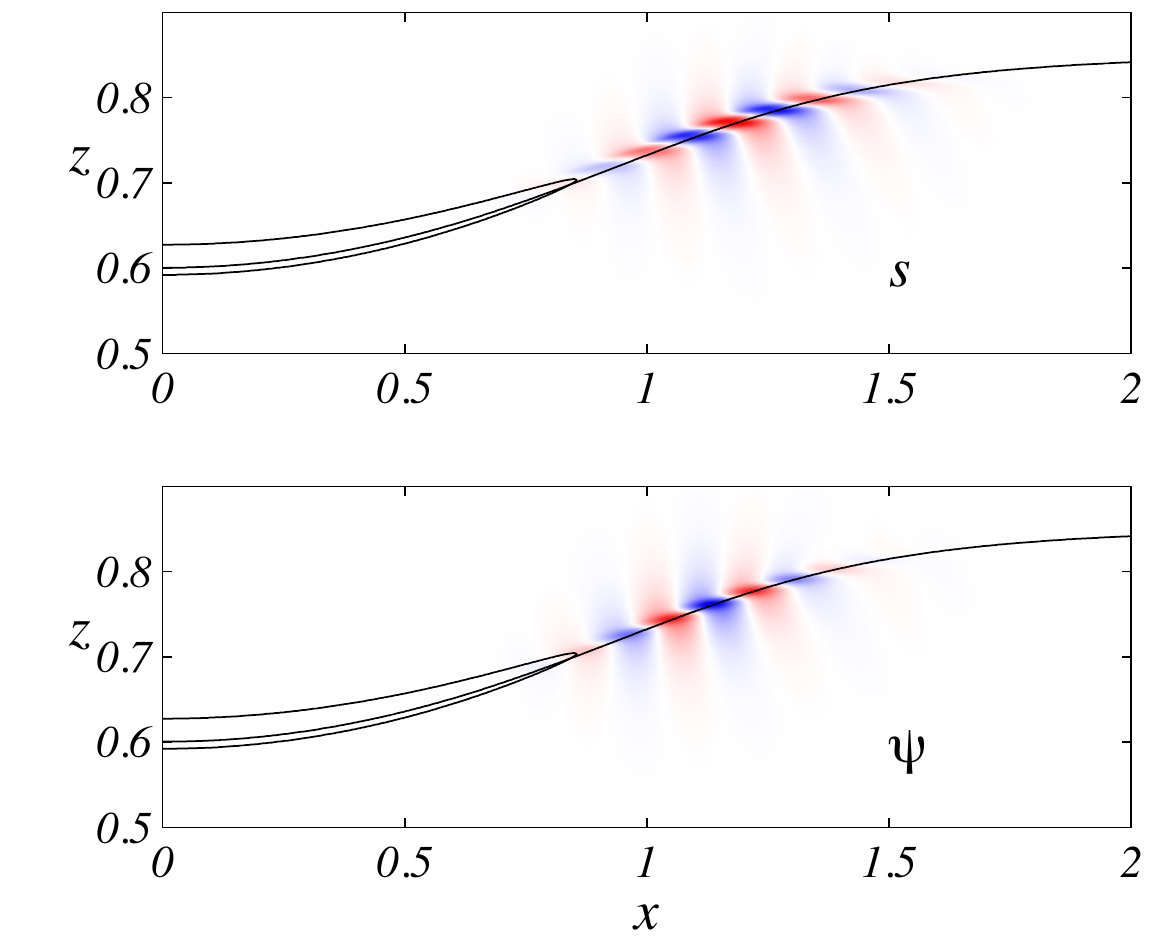}
\put(-80,142){a)}
\hspace{5mm}
\includegraphics[width=60mm]{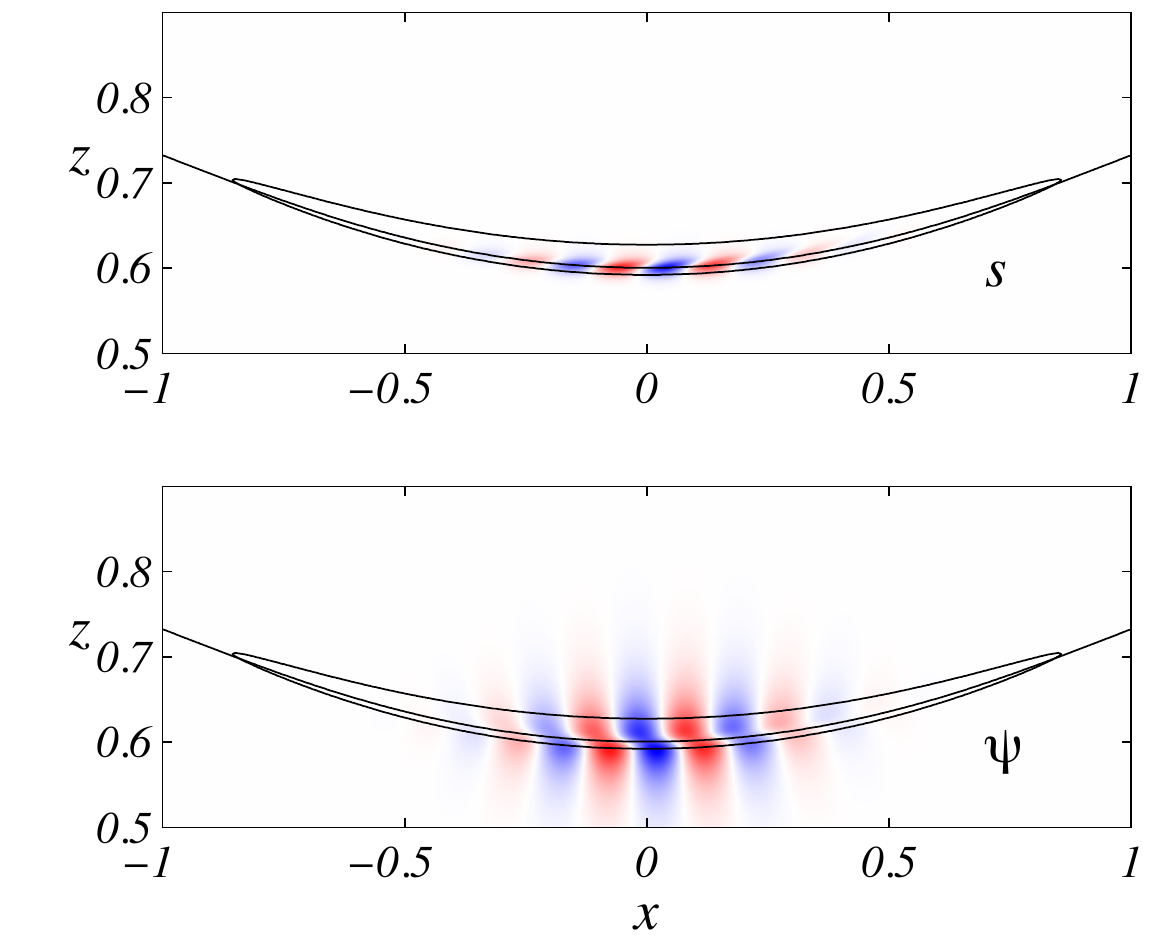}
\put(-80,142){b)}
\caption{a) The structure of the optimal linear perturbation for the $c=0.4810$ DJL wave base state with 
$Re = 10^5$ (at $T=4.43$). The top panel shows the structure of the perturbation density field $s'$ and the lower panel the 
streamfunction $\psi$. The solid lines are the $S=0.5$ and the $Ri = 0.25$ contours. b) The structure of the optimal linear 
at $t = 2.2$ from a forward linear calculation. In both (a) and (b) the perturbation fields have been normalize by their maximum values.}
\label{fig_TOG_structure_c4810}
\end{center}
\end{figure}

The structure and evolution of the optimal perturbation ($T_{opt}=4.43$) at $\Rey = 10^5$ for the $c = 0.4810$ inviscid DJL wave is illustrated in Figure \ref{fig_TOG_structure_c4810}. 
The figure shows the perturbation density $s'$ and stream function $\psi$ (defined by $\psi_z = u'$ and $\psi_x = -w'$) in the frame of a solitary wave that is propagating to the right ($c>0$). The fields have been normalized by their respective maximum values for clarity. The initial ($t=0$) optimal disturbance is a localized wave packet just upstream of the $Ri<0.25$ region (see Figure \ref{fig_TOG_structure_c4810}a). Both $s'$ and $\psi$ are tilted into the ISW-induced shear. The disturbance is dominated by its total kinetic energy which is $\approx 3 \times 10^3$ times larger than the  potential energy. The tilt and dominance of the velocity field over the density field suggests disturbance amplification through the non-normal Orr mechanism \citep{Orr07}. This will be explored further in \S \ref{sec:Orr_growth}.

As the optimal perturbation travels through the ISW, the wave packet is amplified and its structure changes. Figure \ref{fig_TOG_structure_c4810}b shows the packet at $t=2.2$ when it is located at the wave crest. The $\psi$ field is still tilted into the shear, but extends further away from the interfacial region. The $s'$ field is more confined and tilted with the shear. This structure of both fields and the ratio of total potential to kinetic energies, $0.238$, are consistent with a standard unstable K-H normal mode.

\begin{figure}
\begin{center}
\includegraphics[scale=1]{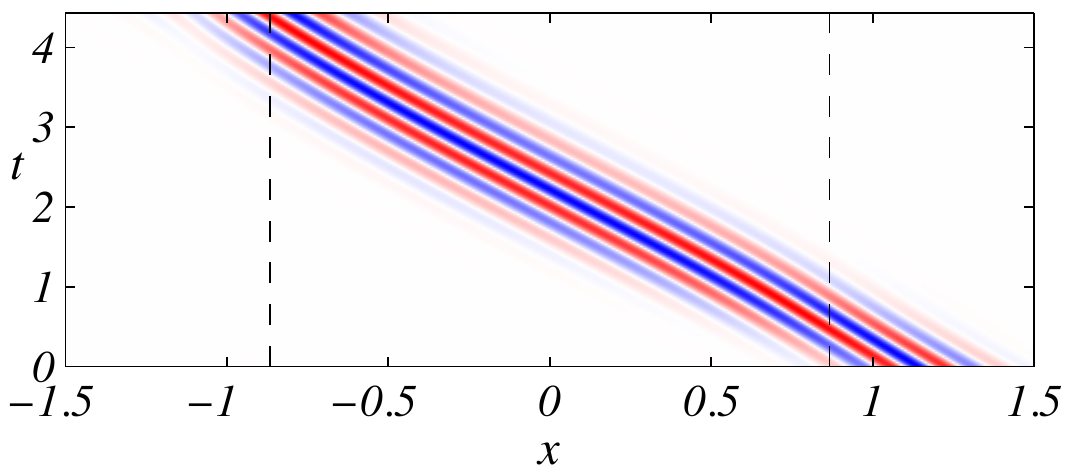}
\caption{An $x$-$t$ plot of $\psi$, normalized to a maximum of one at each time, on the $S=0.5$ contour from the forward linear calculation in Figure \ref{fig_TOG_structure_c4810}(b). The dashed lines are at $|x| = L_{Ri}$. }
\label{fig_psi_on_B05_c4810}
\end{center}
\end{figure}

The temporal evolution of the optimal wave packet further illustrated in Figure \ref{fig_psi_on_B05_c4810} by means of an $x-t$ diagram of $\psi$ along the $S = 0.5$ isopycnal. Again, $\psi$ is scaled by the maximum value at each time. The wave packet remains localized and travels at a quasi-constant group velocity through the solitary wave.  The phase and packet group velocities are nearly identical. The carrier, or central, frequency of the waves in the packet is $\omega = -14.85$. The negative sign is used since $\omega$ is a Doppler shifted value in the frame moving with the solitary wave and we will take all real wave numbers $k>0$. The carrier frequency of the wave packet is found using the discrete Hilbert transform in $t$ of $\psi(x,t)$. At fixed $x$, this gives $Z(t) = H(\psi(t)) = A \exp(i\theta)$, where $H(\cdot)$ is the Hilbert transform. Then $A(t) = |Z(t)|$ is the packet envelop and $d\theta/dt$ is the instantaneous frequency \citep{Oppenheim99}. The carrier frequency is then defined as the average of $d\theta/dt$ for $A/A_0>0.7$, where $A_0 = \max(A)$. These values are then averaged over a range of $x$ around $x=0$ to obtain the reported frequency, $\omega$. Standard Fourier analysis gives estimates generally within $\pm0.1$; however, the Hilbert analysis is used since it also gives the envelop characteristics and allows $\omega$ to be defined in a way that avoids potential dispersive effects at the edges of a packet.  A similar analysis in $x$ at fixed $t$ gives the central wave number of the packet (see the discussion in \S\ref{sec:WKBJ}).  

The structure and behavior of optimal disturbances for all the other ISWs (i.e., $c$) examined are consistent with 
Figures \ref{fig_TOG_structure_c4810} and  \ref{fig_psi_on_B05_c4810}. The optimal packets are initially situated just upstream of the 
$Ri < 0.25$ zone, remain coherent and compact as they move through the ISW, and have a well-defined wave numbers and frequencies. 
As a consequence, a local approach to transient growth might be expected to be relevant. 

\section{Transient growth in the WKB limit}\label{sec:WKBJ}

The present flow geometry appears to be an interesting case for a comparison between the optimal perturbation and the amplification rate predicted by the local linear spatial stability properties of the ISW. In the following, we consider steady-state inviscid DJL waves in the same range, $c=[0.4442, 0.4925]$, as the transient growth analysis in \S \ref{sec:optim}. 
Following \citet{LambF11}, spatial stability analyses for locally parallel ISW base flows were performed to extract the maximum spatial growth rate at each position, from which the perturbation growth in $x$ could be estimated. 

\subsection{Local stability analysis}\label{sec:WKBJ1}

\begin{figure}
\begin{center}
\includegraphics[width=100mm]{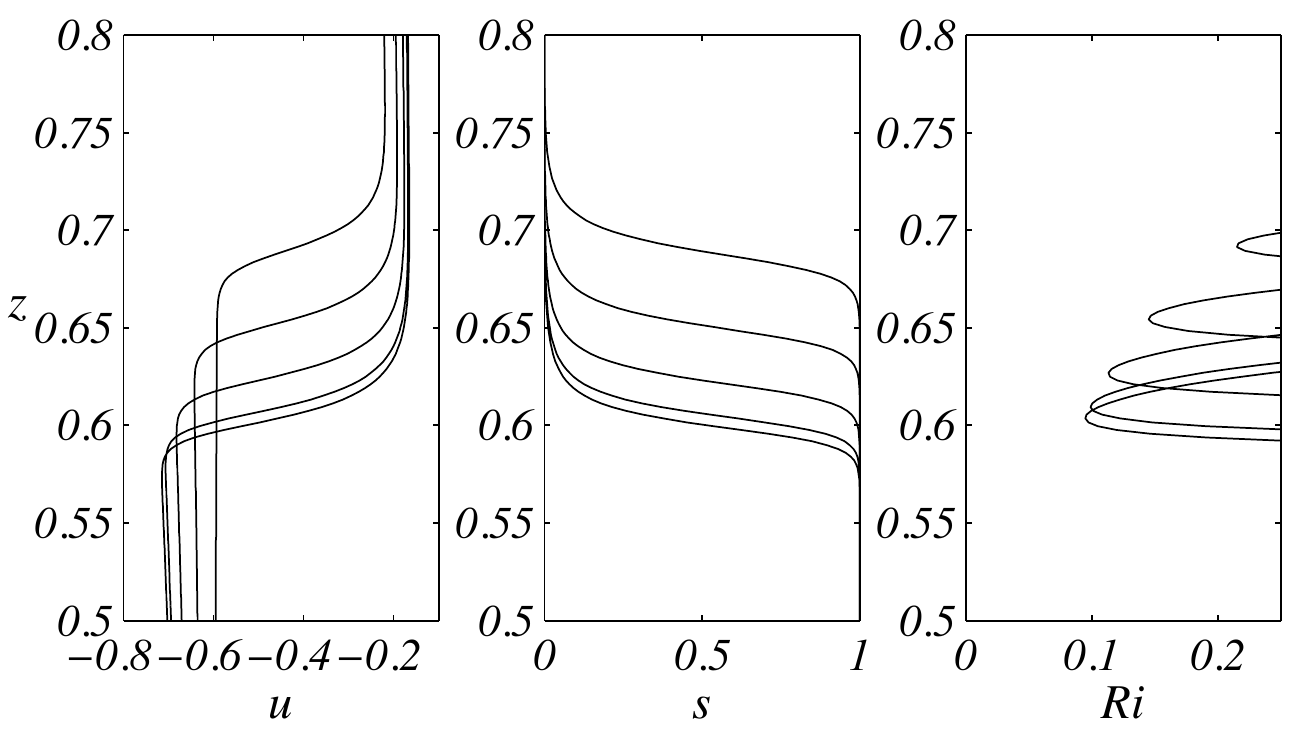}
\put(-245,140){a)}
\put(-150,140){b)}
\put(-60,140){c)}
\caption{Vertical profiles of a) $U(z)$ in the wave frame, b) $S(z)$ and c) $Ri(z)$ for the $c=0.4810$ DJL wave at $|x| = 0,0.2,0.4,0.6$ and $0.8$.}
\label{fig_suRi_c4810}
\end{center}
\end{figure}

Normal modes are sought of the form
\begin{equation}
 \psi = \hat{\psi}(z)\, e^{i(kx-\omega t)},
 \label{phi}
\end{equation}
where the (real) frequency is $\omega$ and the (complex) wavenumber $k=k_r +ik_i$ is the spatial eigenvalue.  Here $k_r$ is the horizontal wavenumber and $k_i$ is the spatial growth rate. 
Linearizing the  Navier-Stokes system for the parallel flow profiles $(U(z),0,S(z))$ shown in Figure \ref{fig_suRi_c4810}(a-b), 
taking the curl of (\ref{direct}) in the $(x-z)$ plane and combining with the density equation, the Taylor-Goldstein (T-G) equation reads
\begin{equation}
\left[\left(\frac{\partial^2}{\partial z^2}-k^2\right) + \frac{k^2N^2 - k(Uk-\omega)U''}{(Uk-\omega)^2}\right]\hat{\psi} = 0,~~~~{\rm where}~~\hat{\psi}(0) = \hat\psi(1)=0,
\label{TG}
\end{equation}
and $N^2(z)=-S_z$ (in non-dimensional variables). The eigenvalue problem (\ref{TG}) is solved for specified background state using MATLAB's {\it bvp5c} adaptive boundary value problem solver \citep{Matlab}.

\begin{figure}
\begin{center}
\includegraphics[width=80mm]{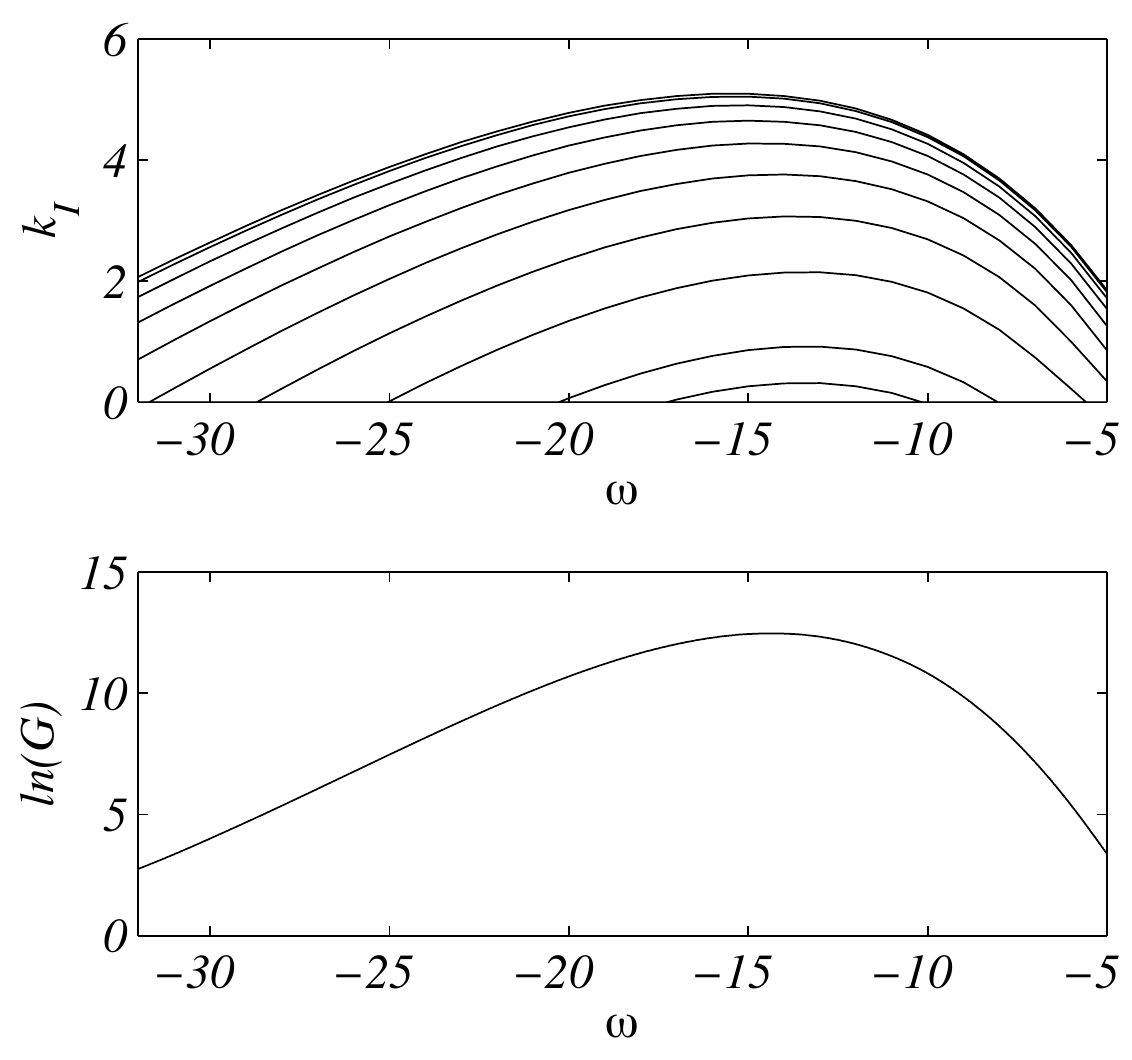}
\put(-180,190){(a)}
\put(-180,85){(b)}
\caption{a) The spatial growth rate $k_i$ versus frequency $\omega$ from the solution of the Taylor-Goldstein equation (\ref{TG})
for the $c=0.4810$ DJL wave. The curves are shown at $|x| = [0$:$0.1$:$0.8]$ from top to bottom. b) The total gain $G$ versus $\omega$ of an unstable perturbation moving through the $Ri<0.25$ zone ($L_{Ri} = 0.846$) from (\ref{Gtot_WKB}) using $k_i(x;\omega)$ in a).}
\label{fig_ki(x)_lnG_vs_omega_c4810}
\end{center}
\end{figure}

Vertical profiles of velocity, density, and Richardson number at several positions within $-L_{Ri}<x<L_{Ri}$ are shown in Figure \ref{fig_suRi_c4810}a for the $c = 0.4810$ ISW. Note that the velocities are in the frame moving with the solitary wave and that the structure is symmetric about $x=0$. As $|x|$ decreases, the minimum Richardson number decreases to $Ri_{min} = 0.096$ at $x=0$. The spatial growth rate $k_i(x;\omega)$ ($>0$ for left-going disturbances) found from (\ref{TG}) with profiles at $|x|=[0$:$0.1$:$0.8]$ for a range of $\omega$ is plotted in Figure \ref{fig_ki(x)_lnG_vs_omega_c4810}a. For fixed $\omega$, the growth rates increase smoothly as $|x| \rightarrow 0$ and the frequency of maximum growth rate,  $\omega \approx -15$, depends only weakly on $x$.

\subsection{WKB approximation to transient growth}\label{sec:WKBJ2}

The local stability properties are linked to the spatio-temporal growth of the perturbation by integrating $k_i(x;\omega)$ from $x=L_{Ri}$ to  $x=-L_{Ri}$.
The evolution of the amplitude $A(x,t)$ of a small amplitude perturbation is therefore given by
\begin{equation}
A(x,t) \sim A(L_{Ri})\exp\left( i \int_{L_{Ri}}^x \left(k^-(x';\omega)-\omega t\right) \,\mbox{d}x'\right).
\label{amp}
\end{equation}
where $k^-$ indicates the waves that grow while propagating to the left ($k_i > 0$) and $A(L_{Ri})$ is the amplitude at the onset of the region $Ri<1/4$. 
The spatial growth of a perturbation with fixed $\omega$ is given by the real part of (\ref{amp}) 
\begin{equation}
\frac{A(x)}{A(L_{Ri})} \sim \exp\left(-\int_{L_{Ri}}^x k_i(x';\omega)\mbox{d}x'\right).
\end{equation}
For these linear disturbances the corresponding energy gain is 
\begin{equation}
G(x;\omega) = \left[\frac{A(x)}{A(L_{Ri})}\right]^2,
\end{equation}
and the total gain after passage of the disturbance through the solitary wave is
\begin{equation}
G(-L_{Ri};\omega) = \left[\frac{A(-L_{Ri})}{A(L_{Ri})}\right]^2.
\label{Gtot_WKB}
\end{equation}

The dependence of the total gain $G$ from (\ref{Gtot_WKB}) as a function of $\omega$ is 
plotted in Figure \ref{fig_ki(x)_lnG_vs_omega_c4810}b for the results in 
Figure \ref{fig_ki(x)_lnG_vs_omega_c4810}a. The maximum gain $\ln(G_{MAX}) = 12.48$ 
occurs at $\omega_{MAX}=-14.74$. Recall that \citet{TroyK05} and \citet{BaradF:10} found observable instabilities required $\bar{\omega}_I T_W > 5$ which gives $\ln(G) > 10$. Similarly, \citet{LambF11} found $2 \bar{k}_i L_{Ri} > 4$, or $\ln(G) > 8$. Figure \ref{fig_ki(x)_lnG_vs_omega_c4810}b shows that $\ln(G) > 8$ occurs over a significant frequency range $-24 < \omega < -7.5$. 

\begin{figure}
\begin{center}
\includegraphics[width=100mm]{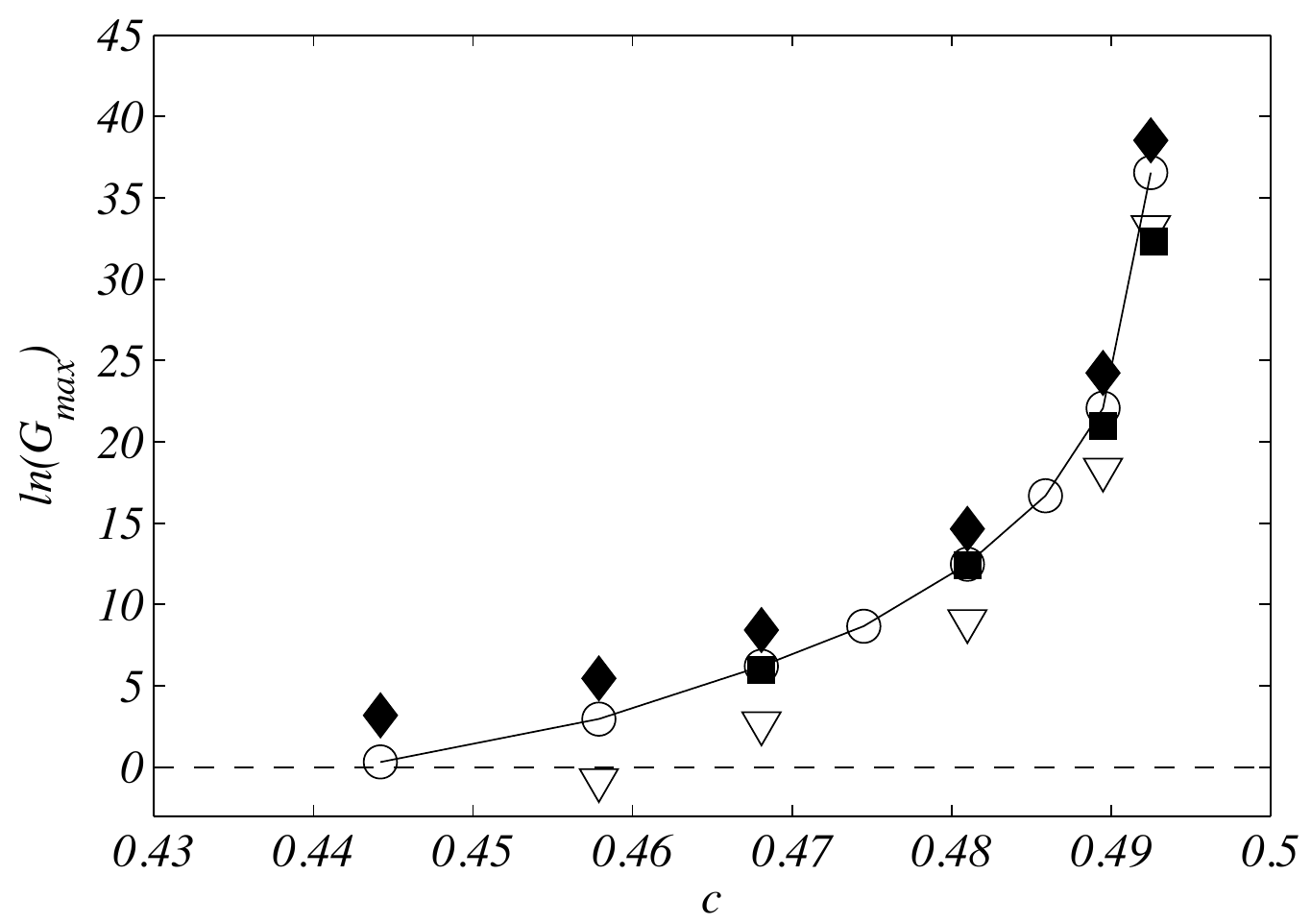}
\caption{The maximum gain, $G_{MAX}$, versus $c$. The solid diamonds (solid squares) are for the DAL optimal perturbations to the inviscid ($Re=10^5$ adjusted) DJL wave base state. The circles are the maximal gains from the spatial WKB analysis. The triangles are for the $k_+$ free wave packets with carrier frequencies equal to those of the optimal disturbances (see \S \ref{sec:free_waves}).}
\label{fig_logGmax_vs_c}
\end{center} 
\end{figure}

Figure \ref{fig_logGmax_vs_c} shows 
a comparison of $\ln(G_{MAX})$ as a function of the ISW phase speed $c$ for the optimal perturbations from the
inviscid DJL  waves, the viscously-adjusted ($\Rey=10^5$) DJL waves, and the WKB analysis described above. The energy gain is always largest for the optimal perturbations on the inviscid DJL wave. Recall that both optimal disturbance calculations were based on $\Rey=10^5$, while the WKB results are for inviscid disturbances. For purely inviscid flows the optimal disturbance growth would be even greater (c.f. Figure \ref{fig_logG_vs_Re}). The difference between the optimal DAL gain and the WKB estimate becomes more evident as $c$, and therefore $L_{Ri}$, decreases. However, the difference in $\ln(G_{MAX})$ is only weekly dependent on $c$, decreasing from $2.9$ to $2.0$ between $c = 0.4442$ and $0.4925$. 
 
\begin{figure}
\begin{center}
\includegraphics[width=80mm]{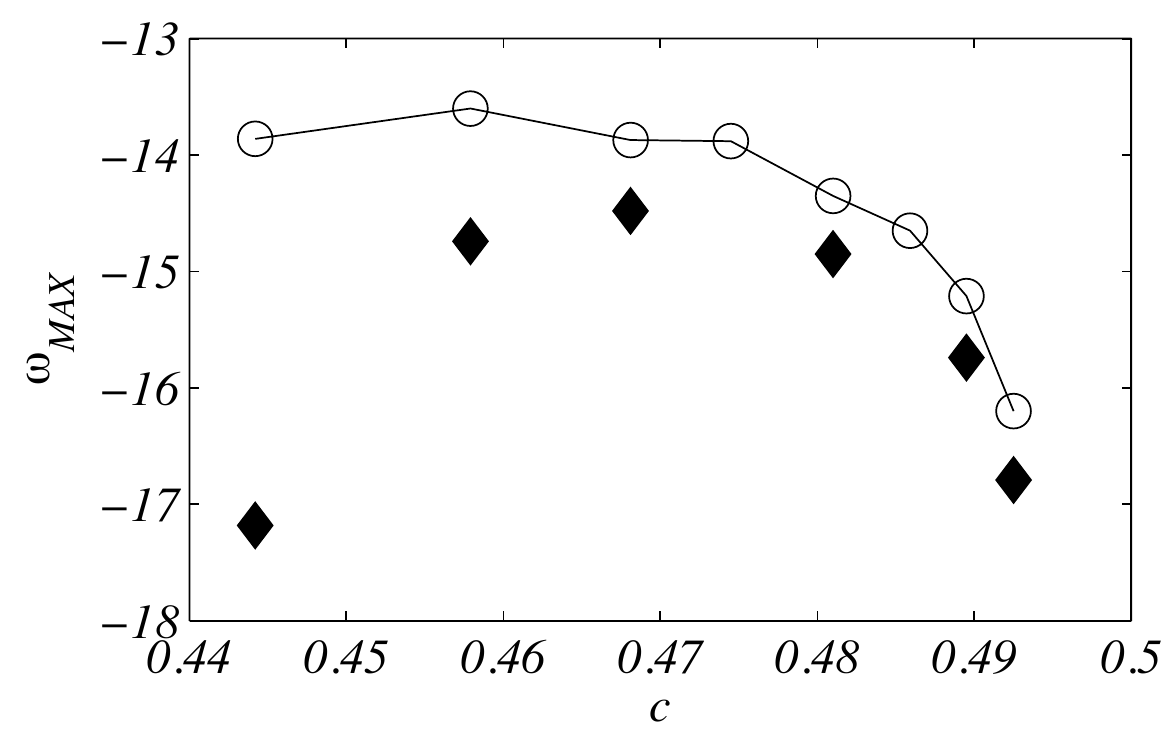}
\caption{The frequency for maximum gain, $\omega_{MAX}$, from the DAL optimal disturbances (solid diamonds) and the spatial WKB analysis (circles) versus $c$.}
\label{fig_omegamax_vs_c}
\end{center}
\end{figure}

The frequencies for maximal WKB growth and the carrier frequencies of the optimal perturbations from the DAL calculation for the inviscid DJL waves are shown in Figure \ref{fig_omegamax_vs_c}. The agreement between the two approaches is similarly quite good, but does degrade as $c$ decreases, where the difference in $\ln(G_{MAX})$ is also greatest.

\begin{figure}
\begin{center}
\includegraphics[width=43mm]{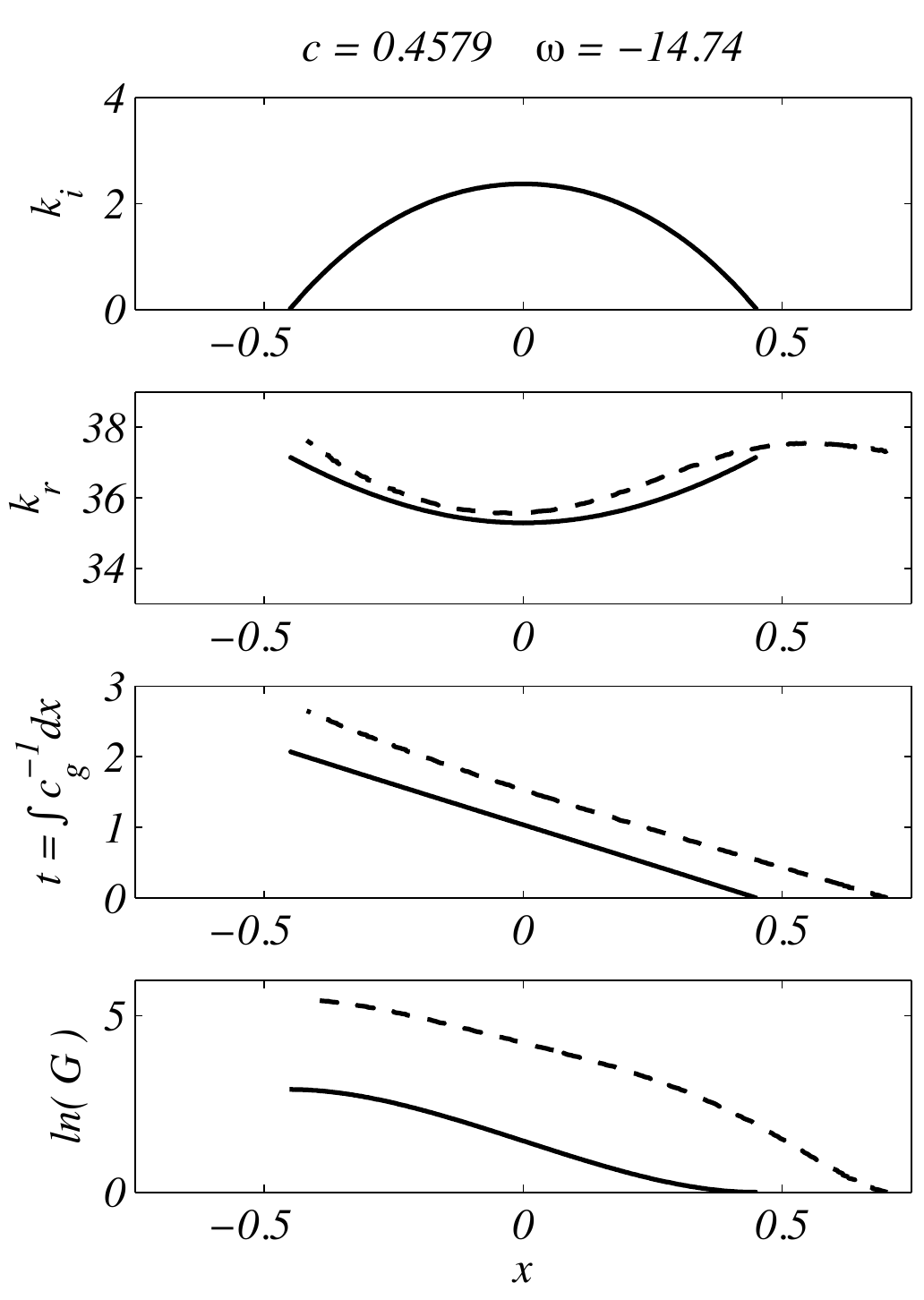}
\put(-120,162){a)}
\hspace{2mm}
\includegraphics[width=43mm]{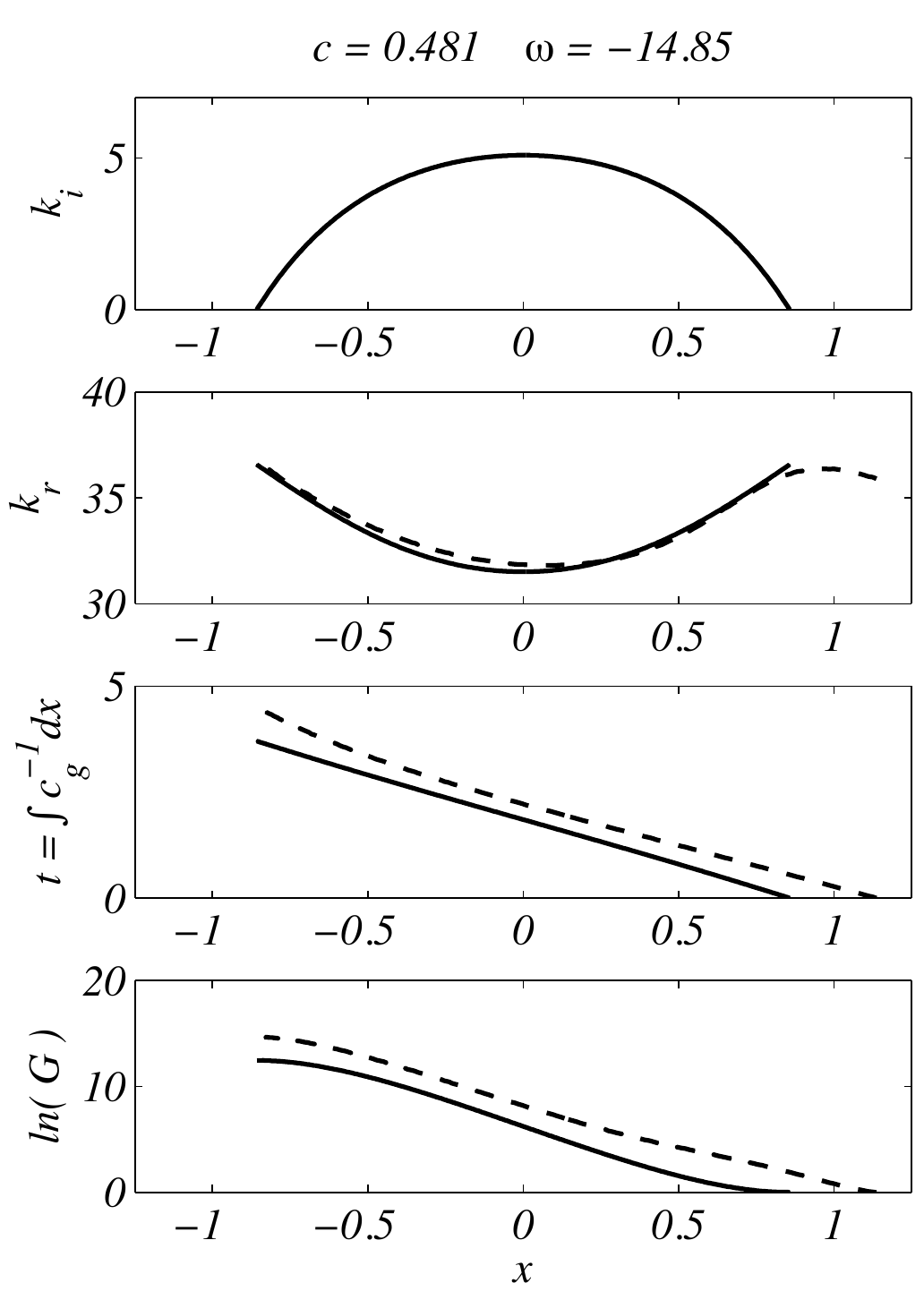}
\put(-120,162){b)}
\hspace{2mm}
\includegraphics[width=43mm]{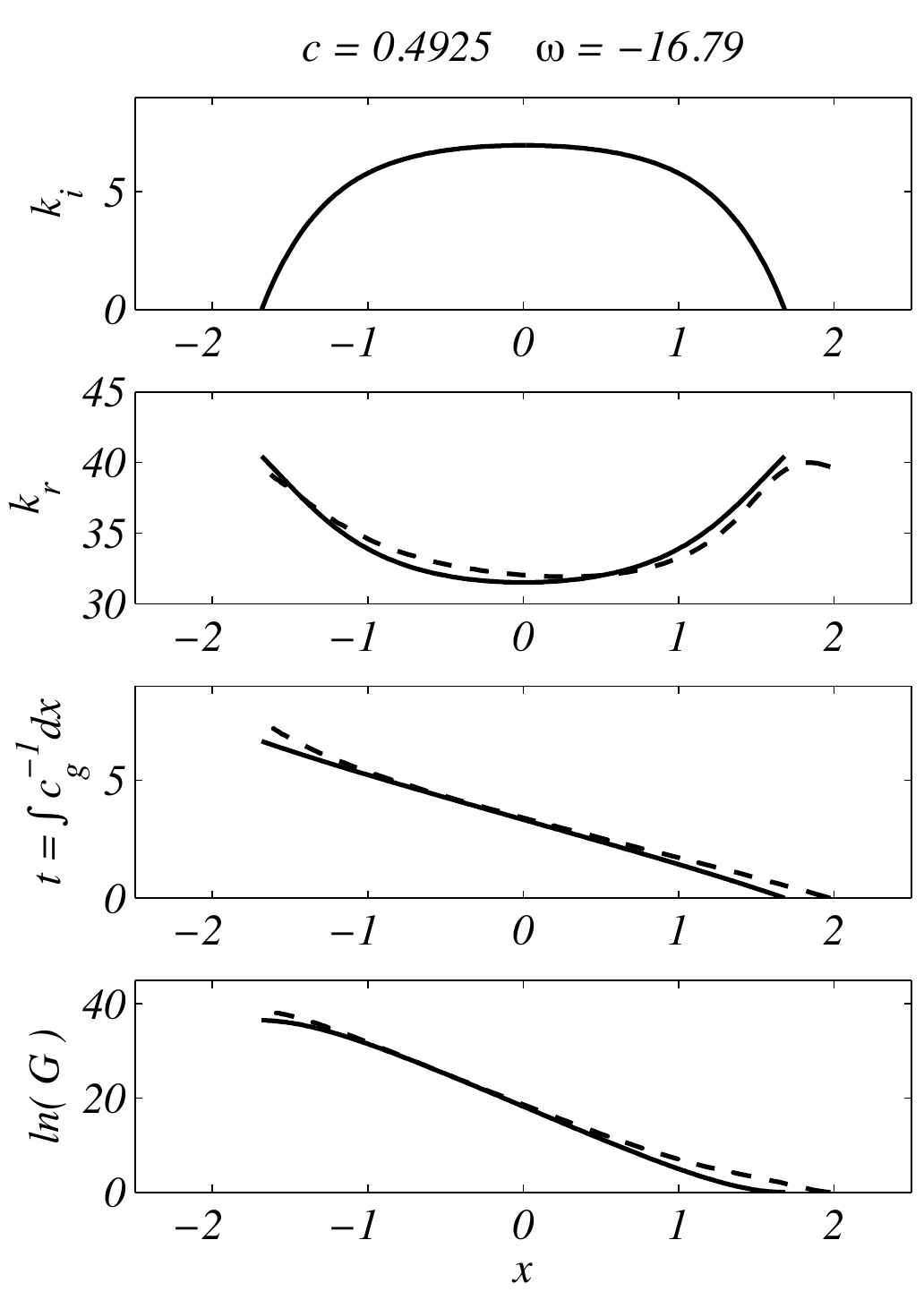}
\put(-120,162){c)}
\caption{$k_i$, $k_r$, $t = \int c_g^{-1} dx$, and $G$ versus $x$ through the $Ri< 0.25$ zones for inviscid DJL waves 
with a) $c = 0.4579$, b) $c = 0.4810$, and c) $c = 0.4925$. The WKB results (solid) are computed for 
$\omega = -14.74$, $-14.85$, and $-16.79$, respectively, the carrier frequencies of the optimal transient growth wave packets 
(see Figure \ref{fig_omegamax_vs_c}). The dashed curves show the corresponding characteristics from forward linear integrations 
of the optimal transient growth disturbances with $Re=10^5$.}
\label{fig_WKB_TOG_k_t_lnG_vs_x}
\end{center}
\end{figure}

A detailed comparison between the disturbance properties through the unstable zone from the linear optimal perturbations, from the DAL method and those from the WKB approach is given in Figure \ref{fig_WKB_TOG_k_t_lnG_vs_x}. The dashed lines in the lower three panels of each column give the wavenumber in the center of the packet, $k_r(x)$, the packet envelop peak position, $x(t)$ (or $t(x)$), and the energy gain, $G(x)$. These were obtained from the forward linear calculations initiated with the optimal DAL disturbances for $c= 0.4579$, $c = 0.4810$ and $c = 0.4925$. As discussed earlier, both $k_r(x)$ and $x(t)$ were found using Hilbert transforms in $x$ of disturbance $\psi(x,t)$ on the $S=0.5$ isopycnal (see Figure \ref{fig_psi_on_B05_c4810}). The solid lines show $k_i(x)$, $k_r(x)$, the disturbance $x-t$ relation,  
\begin{equation}
\nonumber
t(x) = \int_{L_{Ri}}^x c_g^{-1}(s)\mbox{d}s \quad \mbox{ with } \quad c_g(x)= \frac{\partial\omega}{\partial k_r}(x), 
\end{equation}
and $G(x)$ from the WKB approach. The frequency of the optimal perturbations from the DAL analysis (Figure \ref{fig_omegamax_vs_c}) were used in the WKB calculations. The agreement between the central (real) wavenumbers $k_r(x)$ is quite good as is the agreement between the WKB group speed $c_g(x)$ (the slopes of the optimal disturbance trajectories $x(t)$). 
The differences in the energy growth $G(x)$ curves are due almost entirely to an initial growth phase in $x>L_{Ri}$ for the optimal disturbance and  accounts for the difference in $G_{MAX}$ in Figure \ref{fig_logGmax_vs_c}. 
The difference in $\ln(G_{MAX})$ is largely independent of the DJL wave speed $c$, only varying from $2.9$ to $2$ between $c=0.4442$ and $0.4925$. Since vertical shear is present for $x>L_{Ri}$ where $Ri>1/4$, the inviscid Orr mechanism \citep{Orr07} may be responsible for this initial transient growth and is addressed next. 
\subsection{Non-modal transient growth}\label{sec:Orr_growth}

As mentioned above, the WKB analysis does not account for the non-normality associated with the Taylor-Goldstein operator (\ref{TG}). 
In regions where $Ri>1/4$, shear can still play a destabilizing role. In his original work, \citet{Orr07} showed that in the case of a simple inviscid parallel shear flow, perturbations with a non-zero streamwise wavenumber $k$ could produce transient growth through the kinematic deformation of the perturbation vorticity by the baseflow advection and shear.
Later, \citet{FarrellI93_2} derived an analytic solution for the Orr temporal growth rate,
\begin{equation}
\sigma_{Orr} = \frac{1}{2T} \mbox{ ln} \left[ 1 + \frac{U_c'^2 T^2}{2} + U_c'T\sqrt{\left(\frac{U_c'T}{2}\right)^2+1} \right],
\label{Orr}
\end{equation}
for two-dimensional perturbations in a constant, unstratified linear shear $U_c^\prime$ over the 
optimization time $T$. \citet{FarrellI93} also derived an approximation for the Orr gain in the case of an unbounded stratified shear flow where both the shear and the stratification are linear, but in the large $Ri$ limit. Note that in these
idealized cases, $\sigma_{Orr}$ is independent of the streamwise wavenumber $k$.

In the following we examine the possibilities for Orr-type
transient growth in regions of the flow just upstream of the $Ri<0.25$ zone 
using the Taylor-Goldstein equation (\ref{TG}) and maximizing the disturbance energy for short times $T$. Therefore we are seeking a non-modal approach to the optimization problem for the gain
\begin{equation}
G_{Orr}(T) = \max_{\tilde{\mathbf{q}}_0\ne 0}\frac{||\tilde{\mathbf{q}}(T)||^2}{||\tilde{\mathbf{q}}_0||},
\label{Gorr1}
\end{equation}
with $\tilde{\mathbf{q}}=[\psi,s']$. Such problem can be solved considering
the initial value problem associated with the parallel flow Taylor-Goldstein equation (\ref{TG}) using the local buoyancy and velocity profiles. In stream function/density perturbation formulation (\ref{TG}) becomes
\begin{subeqnarray}
\label{TG2}
\Big{[}(\partial_t + ik U)\nabla^2 - ik U'' \Big{]} \psi &=& -ik s',\quad \\
\Big{[}\partial_t + ik U \Big{]} s' &=& ik S' \psi.
\end{subeqnarray}
Recasting (\ref{TG2}) in matrix form 
\begin{equation}
\nonumber
\mathbf{A}\partial_t \tilde{\mathbf{q}} = \mathbf{L} \tilde{\mathbf{q}},
\end{equation}
the initial value problem becomes
\begin{equation}
\tilde{\mathbf{q}}(t) = e^{\mathbf{A}^{-1}\mathbf{L} t}\tilde{\mathbf{q}}_0,
\label{Propag}
\end{equation}
where the initial condition $\tilde{\mathbf{q}}_0=[\psi_0,s_0]$ is to be optimized to maximize (\ref{Gorr1}) at a given time $T$. 
The system (\ref{Propag}) can be solved efficiently by performing a singular value decomposition 
such that \citep{SchmidH01} 
\begin{equation}
G_{Orr}(T)=\sigma_1^2(e^{\mathbf{A}^{-1}\mathbf{L} T}),
\label{Gorr}
\end{equation}
where $\sigma_1$ denotes the first singular value.

\begin{figure}
\vskip4mm
\centering
\includegraphics[width=80mm]{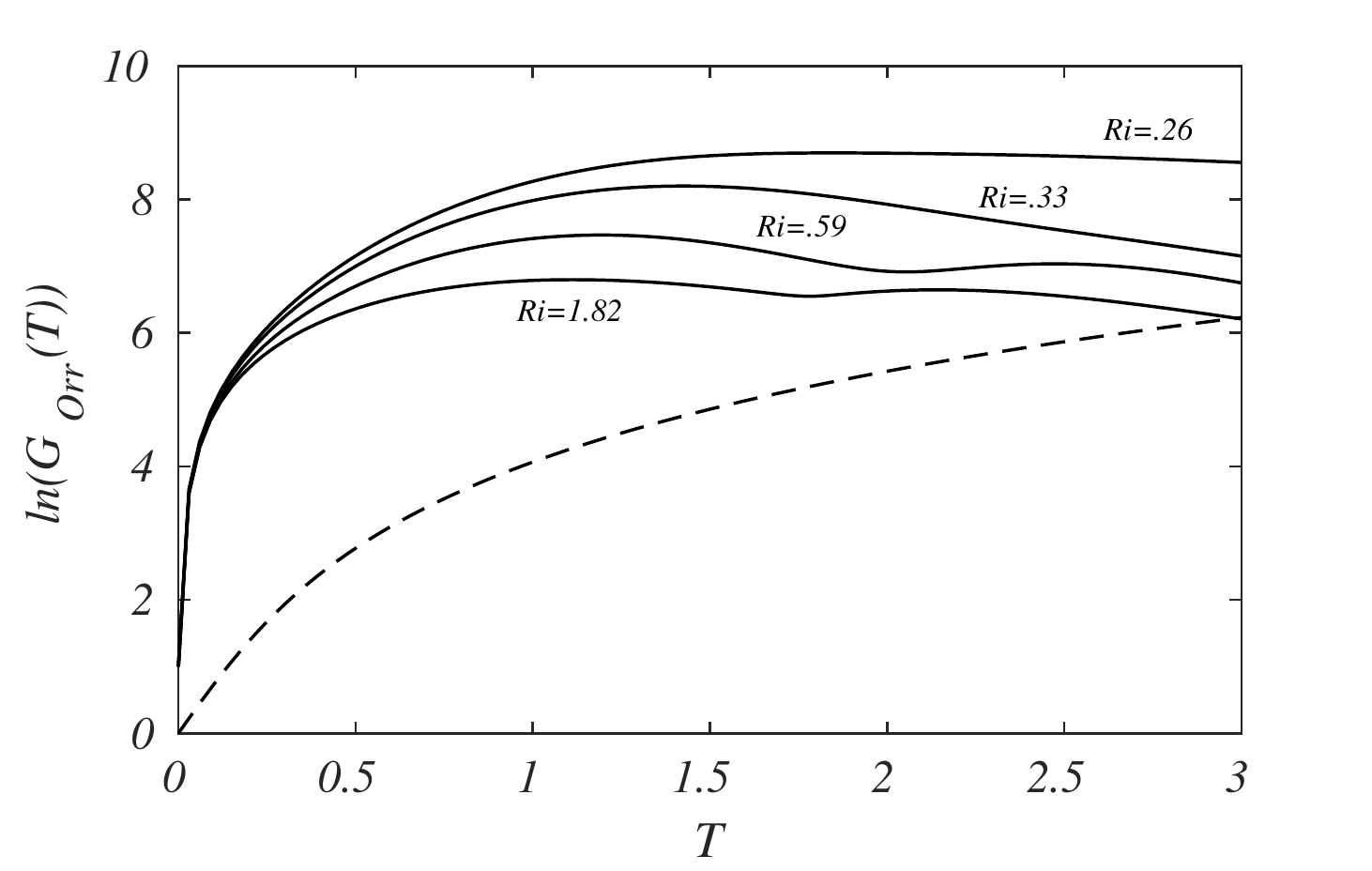}
    \caption{The maximum gain of the Orr mechanism, $\ln(G_{Orr})$, versus the optimization time $T$. The solid lines are the optimal transient growth using the Taylor-Goldstein equation 
              (\ref{TG}) for the stable profiles at $x=[0.88,0.96,1.12,1.36]$ ($Ri_{min}=[0.26,0.33,0.59,1.82]$) for the 
              inviscid DJL wave at $c=0.4810$. The dashed 
              line is the prediction for homogeneous shear from (\ref{Orr}) with $U'_c=7.5$ (i.e. the maximum shear for $Ri_{min}=0.26$).
            }
\label{G_Orr}
\end{figure}

As shown in Figures \ref{fig_logGmax_vs_c} and \ref{fig_WKB_TOG_k_t_lnG_vs_x}, the WKB approach differs by a shift in the energy gain $\ln(G)\approx 2-3$ over an initial advection timescale $t\approx0.5$.
Using the $c=0.4810$ case as an example, the optimal gain (\ref{Gorr}) is computed for the ISW $U(z)$ and $S(z)$ profiles at $x=[0.88,1.12,1,24,1.36]$ which correspond to $Ri_{min}=[0.26,0.33,0.59,1.82]$ respectively. 
The results for $k=36.6$, the average initial packet wavenumber, are shown in Figure \ref{G_Orr}. 
The Orr mechanism contributes to the optimal gain $\ln(G)$ by amounts of  $6-8.5$ and appears to be only weakly affected by changes in $Ri$. The optimal time $T \approx 1$ is twice the time scale in Figure \ref{fig_WKB_TOG_k_t_lnG_vs_x}. However the gain difference between $T=0.5$ and $T=1$ is minor, and for $T>0.5$ the K-H instability provides a faster growth than the Orr mechanism. 

Also included in Figure \ref{G_Orr} is the gain, $\exp({2\sigma_{Orr}T})$, found using (\ref{Orr}) with $U'_c=7.5$. This value is equal to the maximum shear at $x=0.88$. Over times $T < 1$, this simple unstratified model underestimates $\ln(G_{Orr})$ from (\ref{TG2}) and (\ref{Gorr}) by a factor of approximately two. This underlines the important role of stratification in the non-modal growth over short times generally and, in this particular application, before the modal K-H instability dominates.

\section{Linear free wave disturbances}\label{sec:free_waves}

In the limit of linear dynamics the optimal disturbances found above pose the largest possible danger to the internal solitary waves. In the ocean random noise will project on these states so determining the upper bound for linear growth is important in order to provide  bounds on the lifetime of ISWs, as discussed in the following section. ISWs are also subject to encounters with free linear waves propagating on the interface. Indeed disturbances of this sort have been the focus of previous investigations \citep{LambF11,CamassaV12}. Thus it is of interest to explore how these disturbances compare to the optimal perturbations. Additionally, the behavior of these two types of disturbances provides insight into the connections between the optimal perturbation gain, the WKB analysis, the Orr mechanism, and absorption of perturbation energy by the ISW \citep{CamassaV12}. 

\begin{figure}
\begin{center}
\includegraphics[scale=0.6]{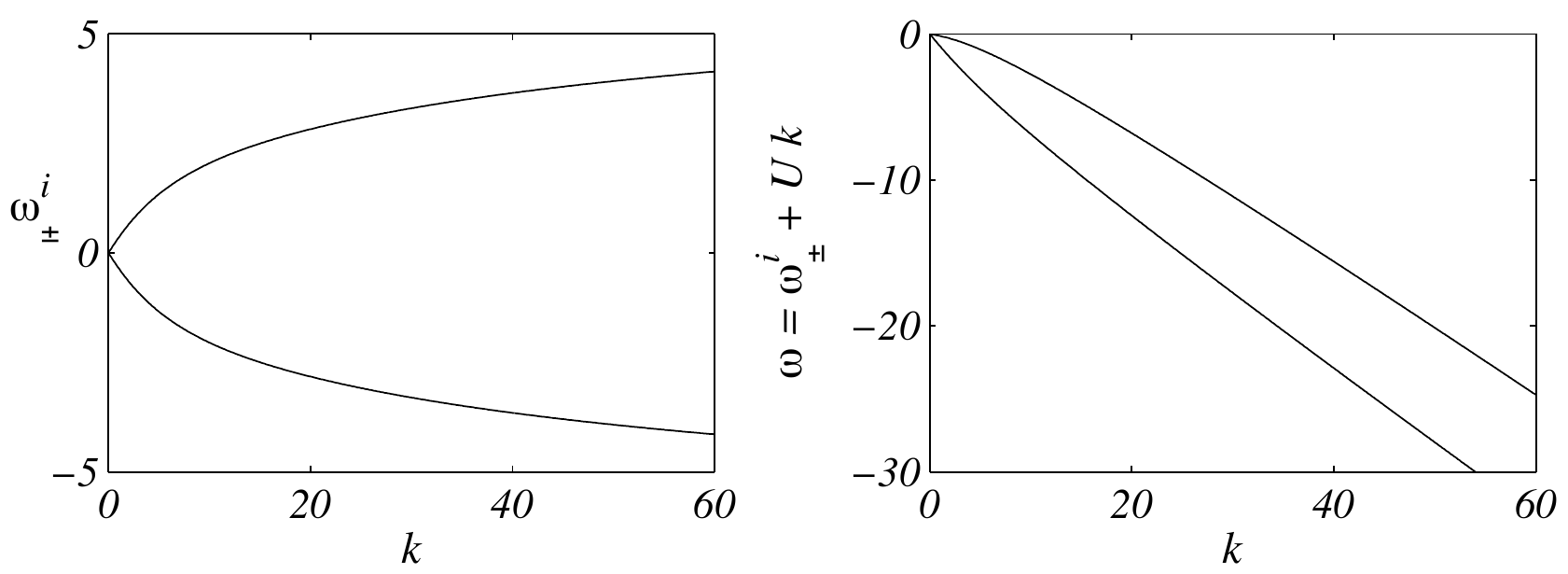}
\put(-220,108){(a)}
\put(-80,108){(b)}
\caption{a) Intrinsic dispersion relation $\omega^i_{\pm}(k)$ for the stationary background stratification. b) Doppler shifted dispersion relation in the ISW frame $U=-c$ and $c=0.4810$.}
\label{disp_rel}
\end{center}
\end{figure}

The characteristics of free linear waves, $\psi=\hat{\psi}(z) \exp[i(kx-\omega t)]$ (for real $\omega$ and $k$), are determined from (\ref{TG}) 
for the undisturbed upstream stratification $\bar{S}(z)$ from (\ref{Sbar}) and no background flow $U=0$. Numerical solution for the first vertical mode wave gives the intrinsic dispersion relation, $\omega^i_{\pm}(k)$, shown in Figure \ref{disp_rel}a for $z_0=0.85$ and $\lambda=80$, and the corresponding eigenfunctions $\hat{\psi}(z)$. The roots correspond to right- and left-going waves propagating toward $\pm x$, respectively. In the ISW frame the frequency, $\omega = \omega^i_{\pm} + Uk$, is Doppler shifted by $U=-c$. Figure \ref{disp_rel}b shows $\omega$ for $c=0.4810$. For a given $\omega$ in the ISW frame there are two waves, designated $k_\pm(\omega)$, corresponding to the right- and left-going roots, respectively. All linear waves have negative phase and group speeds relative to any ISW since $c > c_0$. 

\begin{figure}
\begin{center}
\includegraphics[width=45mm]{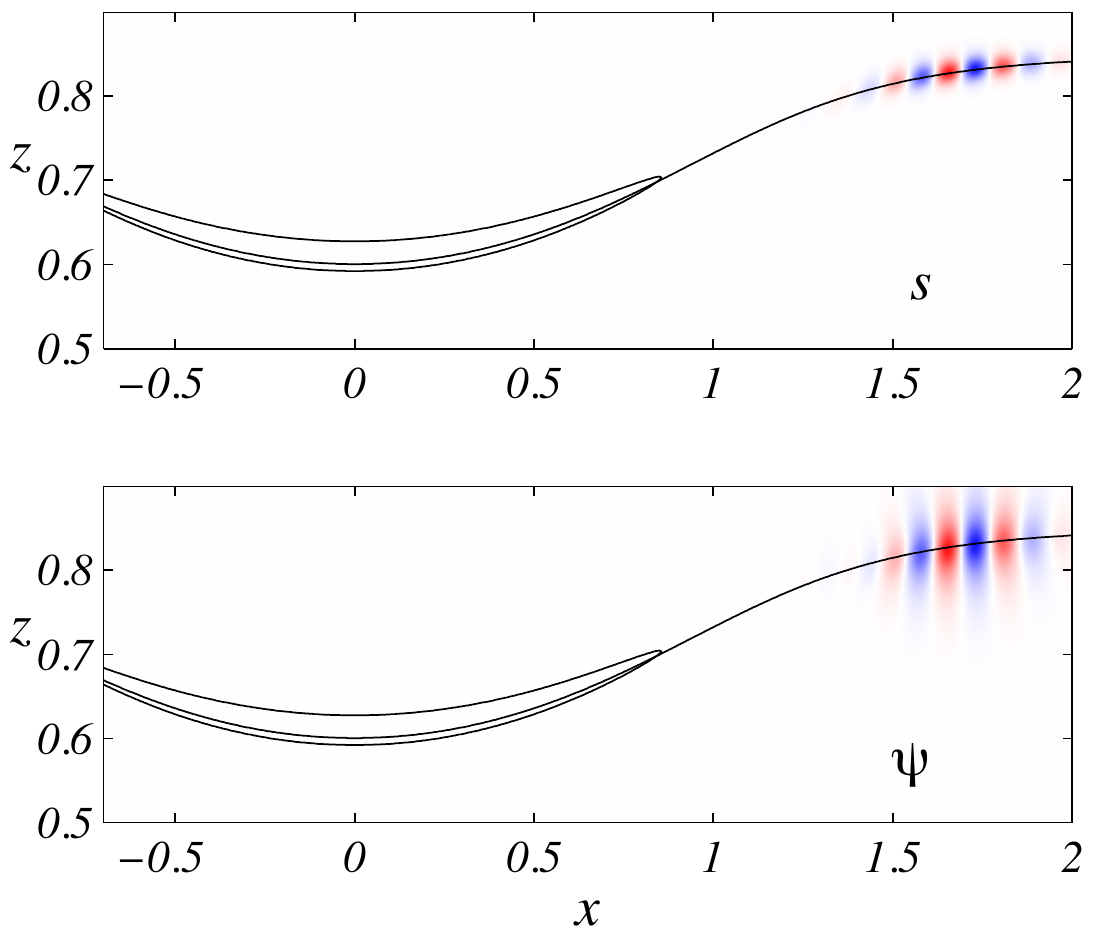}
\put(-65,113){(a)}
\includegraphics[width=45mm]{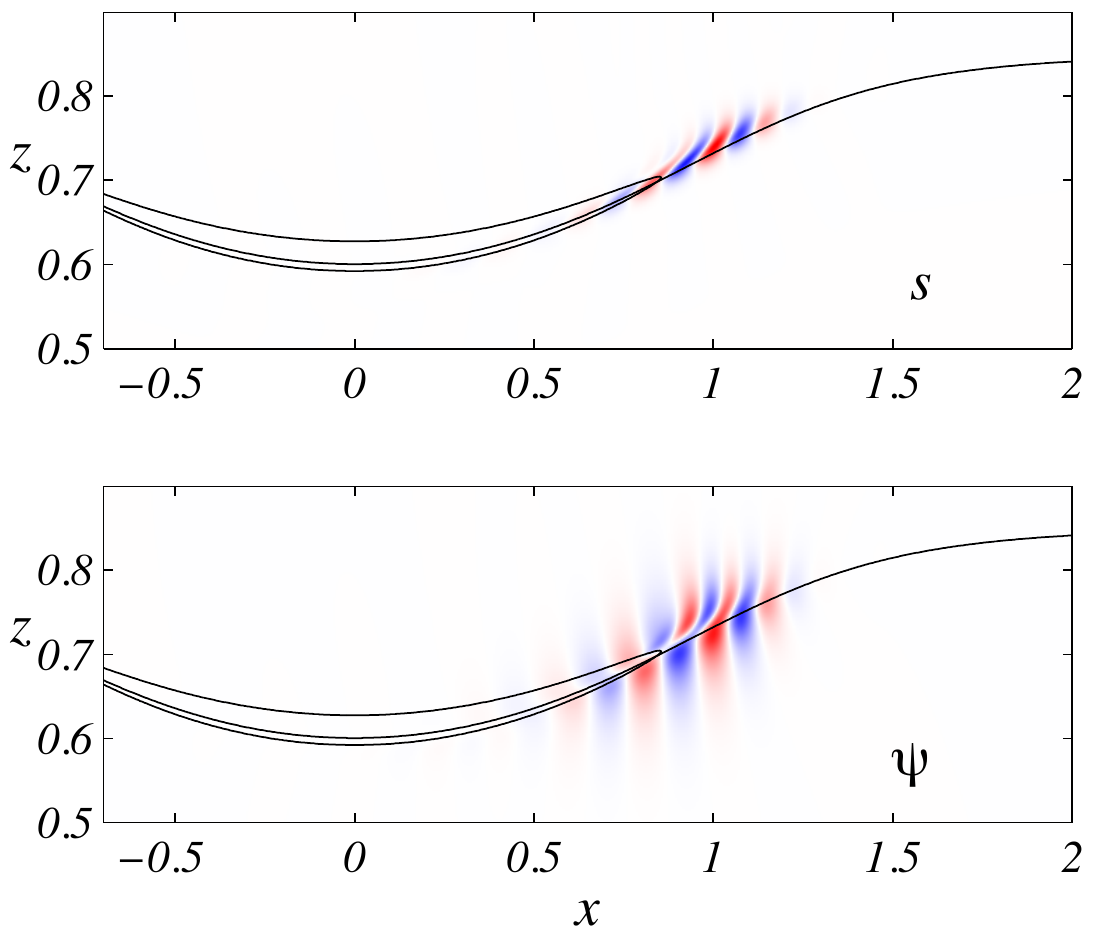}
\put(-65,113){(b)}
\includegraphics[width=45mm]{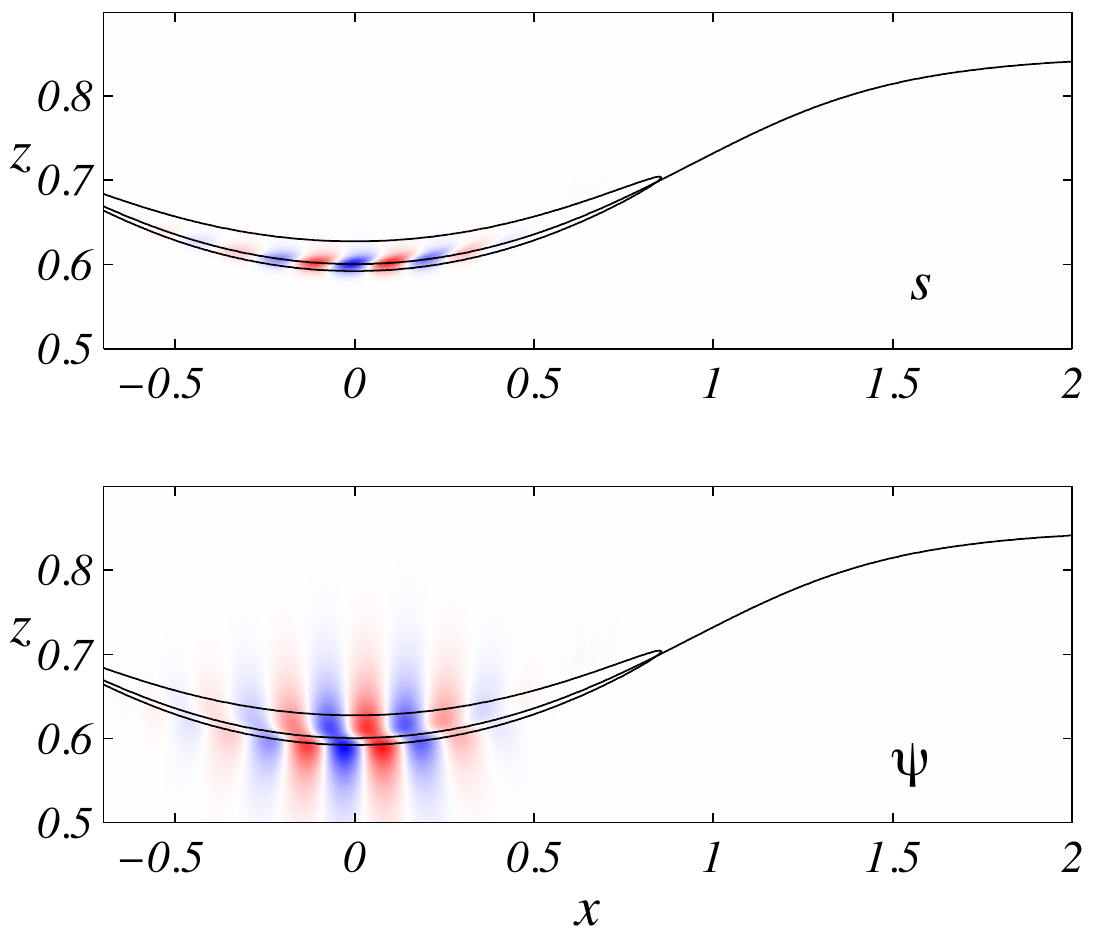}
\put(-65,113){(c)}
\caption{The linear evolution of a Gaussian packet of the right-going waves, $k_+=38.4$, at $\omega = -14.85$ through the $c=0.4810$ DJL wave. The top panels show $s'$ and the lower panels $\psi$. The solid lines are the $S=0.5$ and the $Ri = 0.25$ contours. The fields are shown at a) $t =2.95$, b) $4.9$, and c) $6.55$. Each field has been normalize by its maximum value at that time.}
\label{fig_linwave_FWD_c4810}
\end{center}
\end{figure}

An example of the linear evolution of a Gaussian packet of waves propagating through an ISW is shown in Figure \ref{fig_linwave_FWD_c4810}. The initial condition is 
$$
\psi(x,z,0) = a_0 e^{-\mu^2(x-x_0)^2}\hat{\psi}(z) \cos(kx),
$$
and the companion relation for $s(x,z,0)$. The group is initially centered at $x_0=3$ and the packet width scale $\mu^{-1} = 5$ gives a packet length  approximately equal to an optimal disturbance (see Figure \ref{fig_TOG_structure_c4810}). The amplitude $a_0$ is arbitrary for these linear calculations. In this example the ISW speed $c=0.4810$ and $k_+ = 38.4$ at $\omega=-14.85$, the frequency for the DAL optimal disturbance (see Figure \ref{fig_omegamax_vs_c}). All the linear wave disturbance calculations that follow were made with $\Rey=10^5$ and $Sc=1$ and are the same as used for most of the previous calculations. The packet remains compact and coherent as it enters the ISW (panel a), but just before the unstable zone (panel b) the packet is distorted by the ISW strain field. Once in the unstable zone (panel c) the structure of $s'$ and $\psi$ again closely matches that of an unstable K-H mode. 

\begin{figure}
\begin{center}
\includegraphics[scale=0.75]{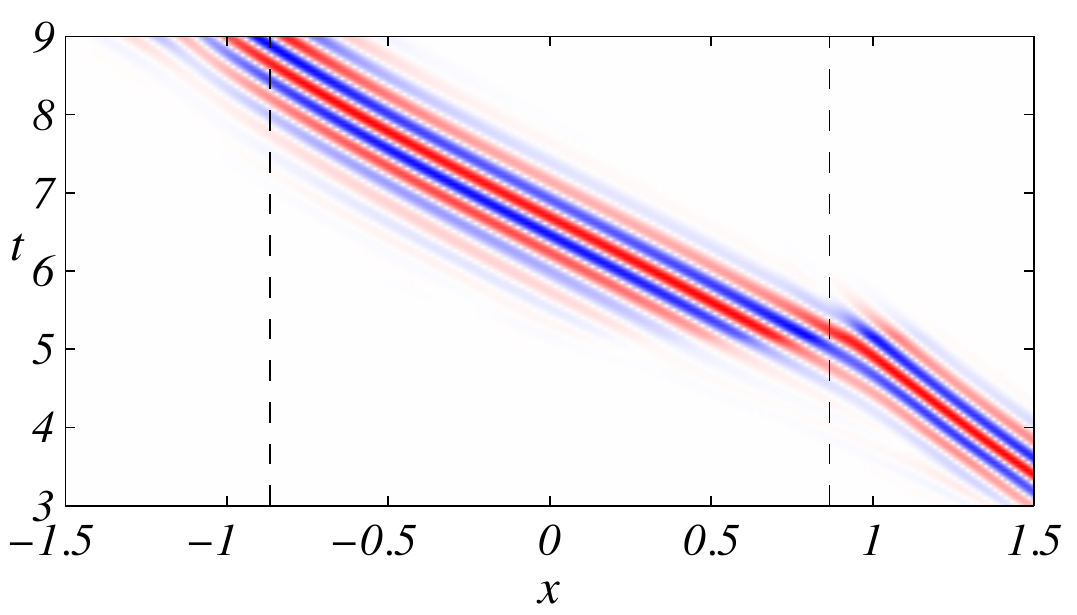}
\caption{$\psi(x,t)$ on the ISW $S=0.5$ contour normalized to a maximum of one at instant from the run in Figure \ref{fig_linwave_FWD_c4810}. The dashed lines are at $|x| = L_{Ri}$.}
\label{fig_linwave_om14p85_psi_xt_B05}
\end{center}
\end{figure}

The evolution of $\psi(x,t)$ on the ISW $S=0.5$ isopycnal (Figure \ref{fig_linwave_om14p85_psi_xt_B05}) further illustrates this scattering and rapid evolution as the packet encounters the leading edge of the unstable zone. In their experimental investigation, \citet{Fructus:09} noted effects of the ISW strain field on disturbances entering the waves. The tilting with the shear of both $s'$ and $\psi$ in Figure \ref{fig_TOG_structure_c4810}b is indicative of transfer of energy from the perturbation field to the ISW \citep{CamassaV12}. It is in contrast to the forward tilt of the DAL optimal disturbance packet in the same region (c.f. Figure \ref{fig_TOG_structure_c4810}a). 

\begin{figure}
\begin{center}
\includegraphics[width=80mm]{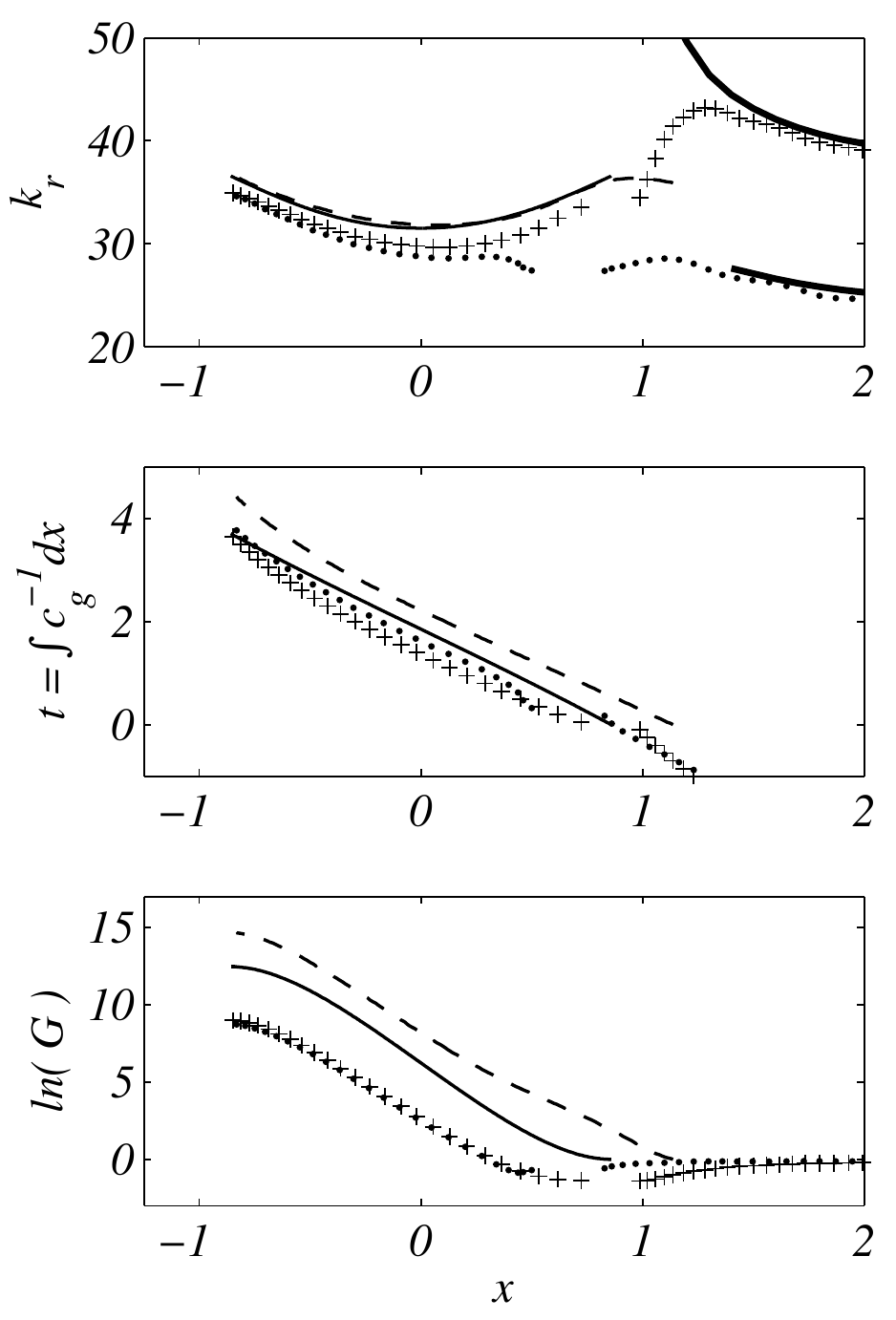}
\caption{The $k_r(x)$, $t(x)$, and $\ln(G(x))$ panels of  Figure \ref{fig_WKB_TOG_k_t_lnG_vs_x}b with the addition of results from the forward linear calculations initiated with  Gaussian-shaped wave packets at $x=3$ with $\omega=-14.85$ and $k_+=38.4$ ($+$) and $k_-=24.6$ ($\bullet$). The heavy solid lines in the top panel show $k_\pm(x)$  calculated from (\ref{TG}) with the stable upstream ISW background flow.}
\label{fig_WKB_TOG_FWD_k_t_lnG_vs_x_c4810}
\end{center}
\end{figure}

The spatial evolution of the packet central wavenumber $k_+(x)$, peak location $x(t)$, and energy gain $G(x)$ are shown ($+$ symbols) in Figure \ref{fig_WKB_TOG_FWD_k_t_lnG_vs_x_c4810} along with the results from the DAL optimal disturbance and the WKB analysis (repeated from Figure \ref{fig_WKB_TOG_k_t_lnG_vs_x}b). Also shown by the thick solid line is $k_+(x)$ found from local, parallel solutions of (\ref{TG}) using the $U$ and $S$ fields from the $c=0.4810$ ISW. As the linear waves approach $x=L_{Ri}$ ($=0.867$), $k_+(x)$ increases rapidly and is discontinuous with $k_r(x)$ at $x=L_{Ri}$. The wave packet in the forward linear calculation closely follows this prediction as it first enters the ISW, but the rapid variation of the background flow and the resultant strong distortion of the packet
causes the curves to diverge. Once in the unstable zone the central wavenumber of the linear packet is slightly smaller than the WKB and DAL optimal results, but does evolve similarly. The packet peak location $x(t)$ decelerates on entering the $Ri<0.25$ region, then accelerates (jumps) as the packet reforms in the unstable zone to approximately the same group speed as the WKB and DAL optimal cases. Note that the time origin has been shifted to $t=0$ when the packet peak is at $x=L_{Ri}$. 
The evolution of the gain $G(x)$ shows an initial loss phase as the packet enters the ISW and then a rapid growth just as found by \citet{CamassaV12}. While still large, the total gain, $\ln(G)=8.75$ is well below the values of $12.44$ and $14.68$ from the WKB analysis and DAL optimal disturbances, respectively.

Both \citet{LambF11} and \citet{CamassaV12} found that the shorter $k_+$ wave to have larger growth through the $Ri<0.25$ zone. However, the left-going wave with $k_-=24.6$ ($\omega=-14.85$), included in Figure \ref{fig_WKB_TOG_FWD_k_t_lnG_vs_x_c4810} (the $\bullet$ symbols),  experiences slightly more total growth, $\ln(G)=8.99$, than the $k_+$ wave. This wave packet does not experience as much initial energy loss entering  the ISW as the $k_+$ wave. Once in the unstable zone, the packet central wavenumber approaches the result of the $k_+$ packet (top panel). The calculation for $k_-(x)$ is stopped at $x\approx1.4$ after a critical layer, $U(z)-c=0$, at $z=0$ appears. Additional calculations for ISWs with $c=0.4895$ and $0.4925$ at $\omega_{MAX}$ from the DAL optimal perturbations and for $c=0.4810$ with $-24<\omega<-7.5$ produced nearly the same total energy gain for the $k_-$ and $k_+$ packets. The reasons for this difference from the previous investigations is not clear. However, these calculations were done for perturbation dynamics linearized about an internal solitary wave, while \citet{LambF11} explored finite-amplitude growth in a fully nonlinear model. \citet{CamassaV12} also employed a nonlinear model, but did try to insure that their disturbances remained in the linear regime. The finite viscosity could be important since more damping of the shorter $k_+$ wave is to be expected. However, calculations at $\Rey = 5\times 10^7$ give identical behavior to that shown in Figure \ref{fig_WKB_TOG_FWD_k_t_lnG_vs_x_c4810} with the exception of more total energy growth ($\ln(G) = 9.98$/$9.77$ for $k_+$/$k_-$).

The total gain of the $k_+$ wave packets with frequencies equal to optimal disturbance $\omega_{MAX}$ and $\Rey=10^5$ are shown as the open triangles in Figure \ref{fig_logGmax_vs_c}. Growing disturbances require $c \gtrsim 0.46$ and the total energy gain is always below both the WKB and DAL optimal disturbance values. Interestingly, the difference in gain between the optimal disturbances and the free wave packets, $\ln(G_{opt})-\ln(G_{fwp})$, is almost  constant, varying only from $6.25$ at $c=0.4579$, where the linear packet experiences a net loss of energy, to $5.35$ at $c=0.4925$. (The difference $\ln(G_{opt})-\ln(G_{WKB})$ varied from $2.9$ to $2$ over the same range.) This difference is comparable to the gain attributable to the Orr mechanism. Recall that for $c=0.4810$ growth of $\ln(G)\approx 6-7$ is possible over a time period of $t \le 0.5$ when a disturbance is in the strongly sheared, but stable, region just upstream of $x=L_{Ri}$ (see Figure \ref{G_Orr}). The time for a free wave packet to propagate the same distance is $t \approx 0.6$. This implies that the difference between the linear growth of a free wave packet and the DAL optimal disturbance is almost entirely due to the non-normal effects. The optimal disturbances are structured to utilize the Orr mechanism, while the free waves first loose energy to the ISW, also through non-normal dynamics \citep{CamassaV12}, before the absorption is reversed through the K-H mechanism. The WKB result falls between these two and does not involve either of these non-normal influences. 

We attempted to quantify the contribution of both, the Orr and the absorption mechanism and performed a similar energy budget to \citet{CamassaV12} during the early development of the DAL optimal perturbation. However, such analysis does not allow for splitting the contributions of each mechanism in the energy budget. Nevertheless, it is interesting to note that the Orr gain $\ln(G_{Orr})$ roughly corresponds to the difference between the DAL optimal perturbation gain $\ln(G_{opt})$ and the gain produced by the packet of linear free waves $\ln(G_{fwp})$, at least for the DJL waves considered in the present study.

\section{Nonlinear evolution}\label{sec:nonlin}

Several previous investigations have explored the nonlinear evolution of disturbances on thin-interface internal solitary waves, with \citet{Almgren:12}, and \citet{LambF11} being the most relevant to this discussion. However, both of these studies focussed on free wave disturbances, while here the finite-amplitude evolution of DAL optimal disturbances is of interest. To maintain consistency with these earlier studies the finite-volume, incompressible Navier-Stokes code, IAMR \citep{Almgren:98}, is used. This model is an adaptive-grid version of the VARDEN code used in \citet{Almgren:12} and \citet{CamassaV12} and the models in \citet{BaradF:10} and \citet{LambF11} are built on the same underlying second-order advection and projection algorithm. 

The present calculations are made for two-dimensional flow on a fixed, isotropic grid with cells sizes $\Delta x=\Delta z=1/512$ and a rigid lid. The runs were initiated with a DJL solitary wave centered in the domain of half-length $L\ge6$ along with any initial perturbations. The upstream boundary, $x=L$, was an inflow boundary with $s=\bar{S}(z)$ from (\ref{Sbar}) and $U = -c$. Since $c>c_0$, disturbances are advected out of the open boundary at $x=-L$. The open boundaries require that IAMR be run in a non-Boussinesq mode, therefore all the results presented below use $\Delta \rho /\rho_0 = 0.01$ so that the flow will be close to Boussinesq and analyzed as such. The calculations have zero explicit viscosity and diffusivity. The Godunov-based advection scheme produces an implicit numerical diffusion, but it only becomes significant when gradients at the grid scale are large. 

An important diagnostic is the (non-dimensional) domain integrated Boussinesq ISW energy per unit width $E_{ISW}=E_{KE}+E_{APE}$, where 
$$E_{KE}=\frac{1}{2}\int_\Omega  |\mathbf{u}(\mathbf{x},t)|^2 \mbox{d}\Omega$$ and 
$$
E_{APE}=\int_\Omega \int_{\bar{S}(z)}^{s(\mathbf{x},t)} \left( z - \bar{z}(s')\right) \mbox{d}s'\mbox{d}\Omega,
$$ 
are the total kinetic and the available potential energies, respectively. The $E_{APE}$ is found using the background density field $\bar{S}(z)$ appropriate for ISWs \citep{Scotti:06,Lamb:08} and $\bar{z}$ is the inverse mapping of $\bar{S}(z)$ such that $\bar{z}(\bar{S}(z,t))=z$.

\begin{figure}
\begin{center}
\includegraphics[width=65mm]{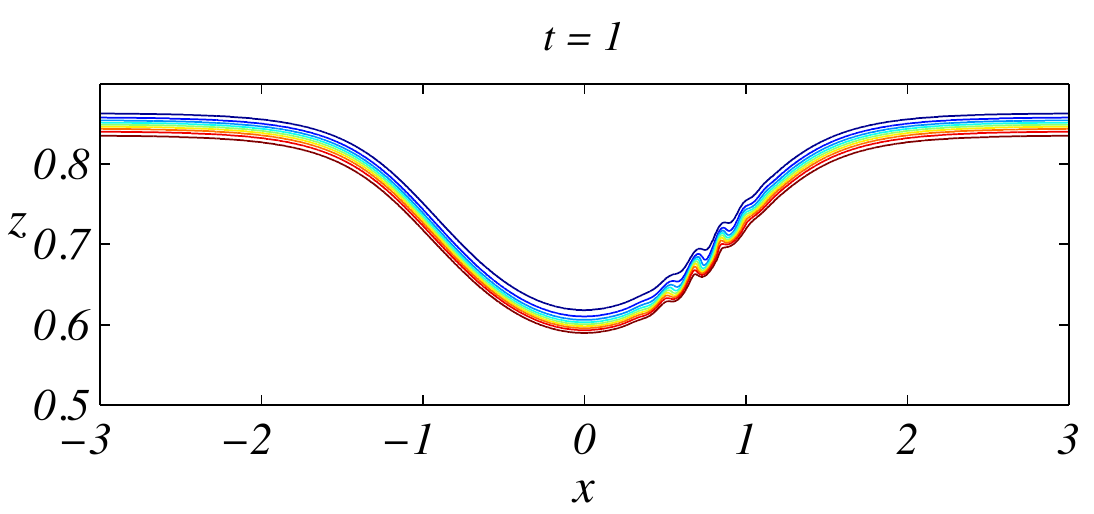}\put(-155,35){(a)}\includegraphics[width=65mm]{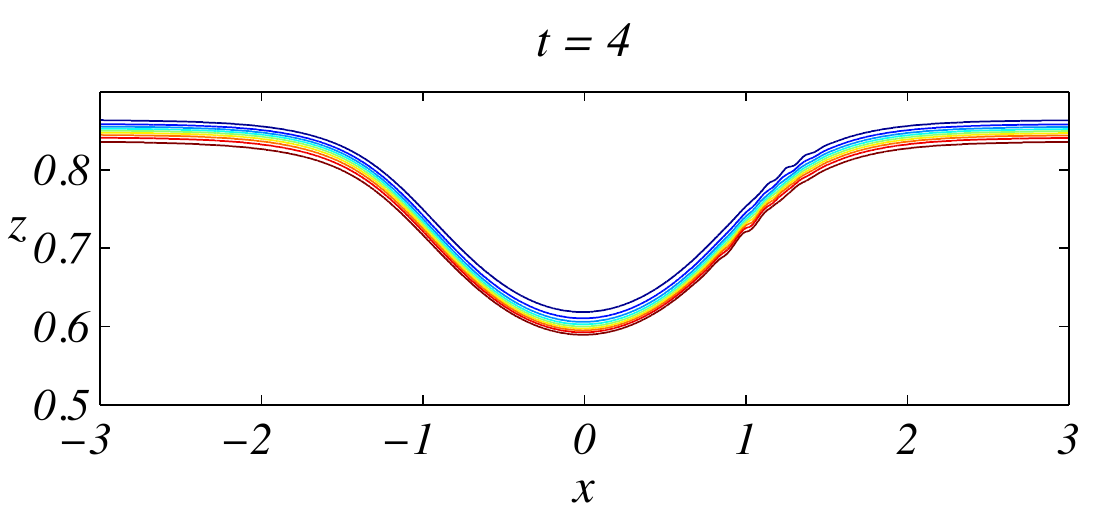}\put(-155,35){(f)}\\ \vspace{3mm}
\includegraphics[width=65mm]{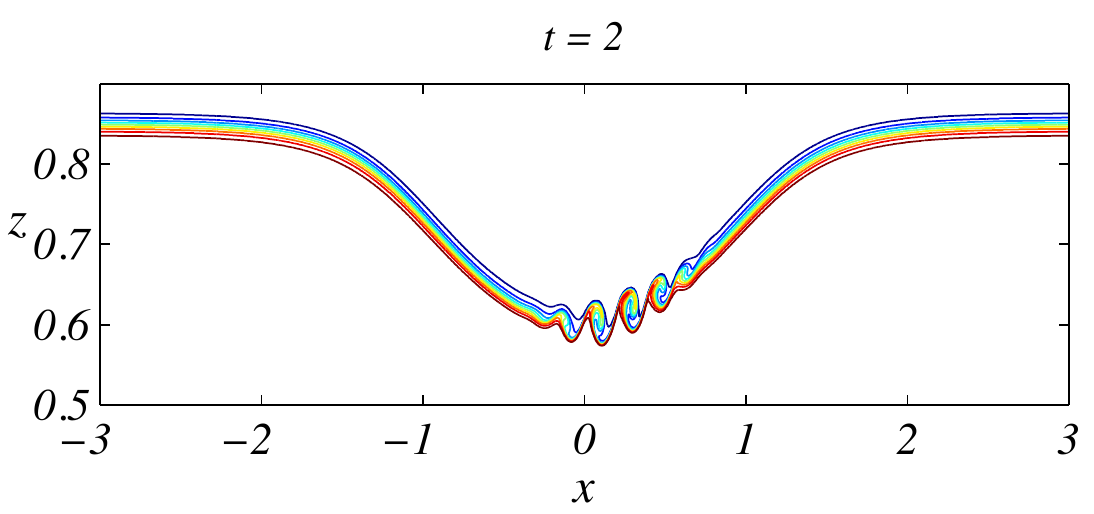}\put(-155,35){(b)}\includegraphics[width=65mm]{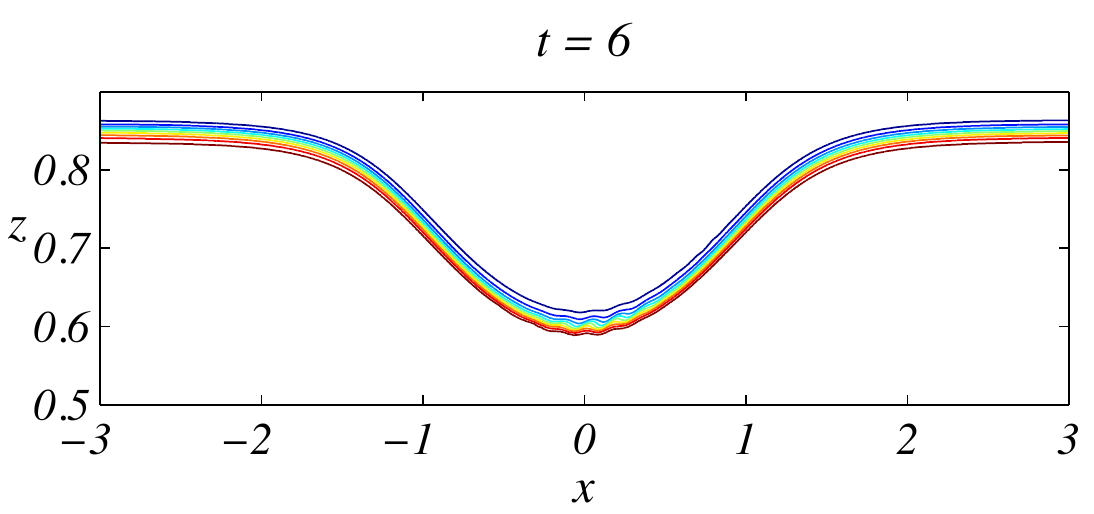}\put(-155,35){(g)}\\ \vspace{3mm}
\includegraphics[width=65mm]{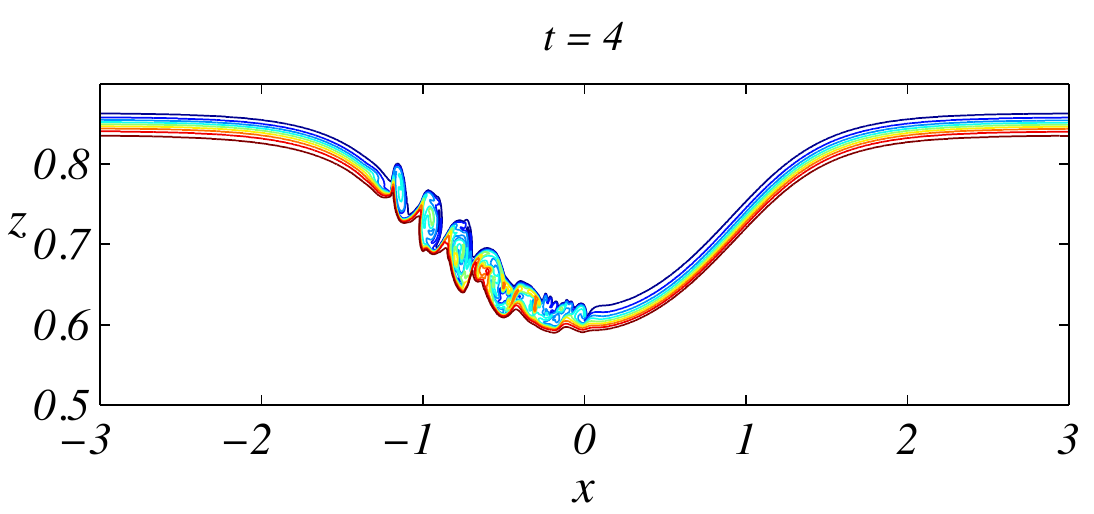}\put(-155,35){(c)}\includegraphics[width=65mm]{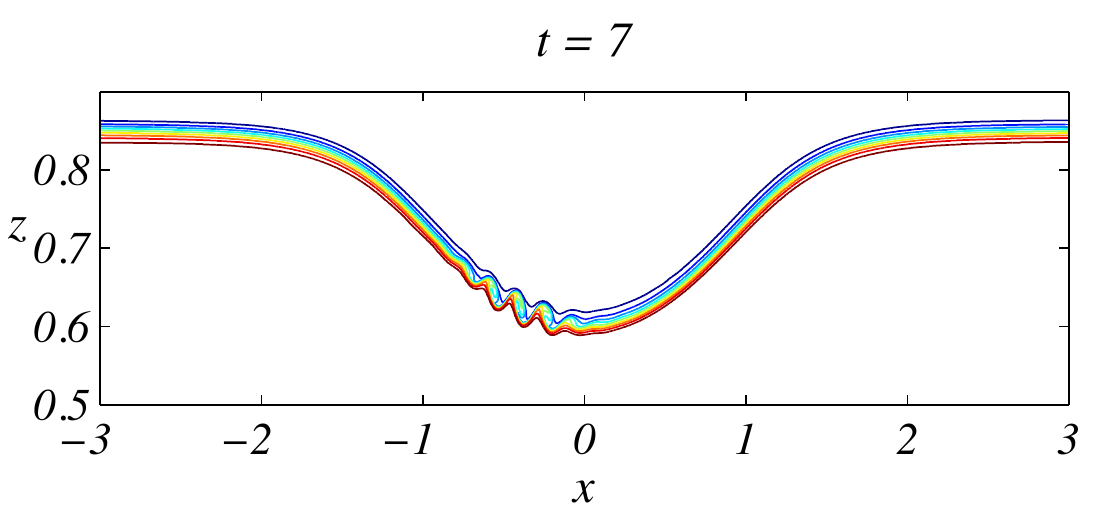}\put(-155,35){(h)}\\ \vspace{3mm}
\includegraphics[width=65mm]{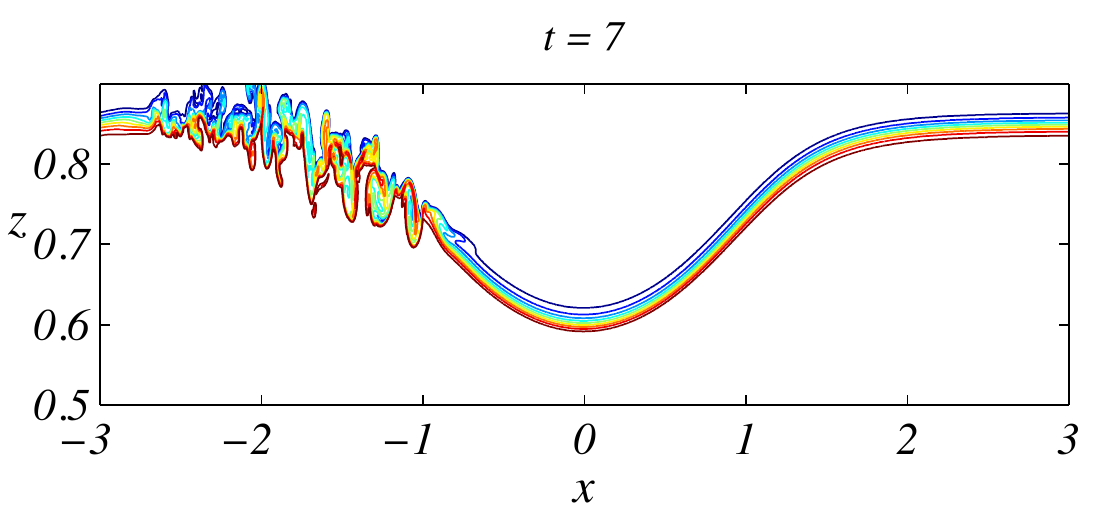}(\put(-155,35){(d)}\includegraphics[width=65mm]{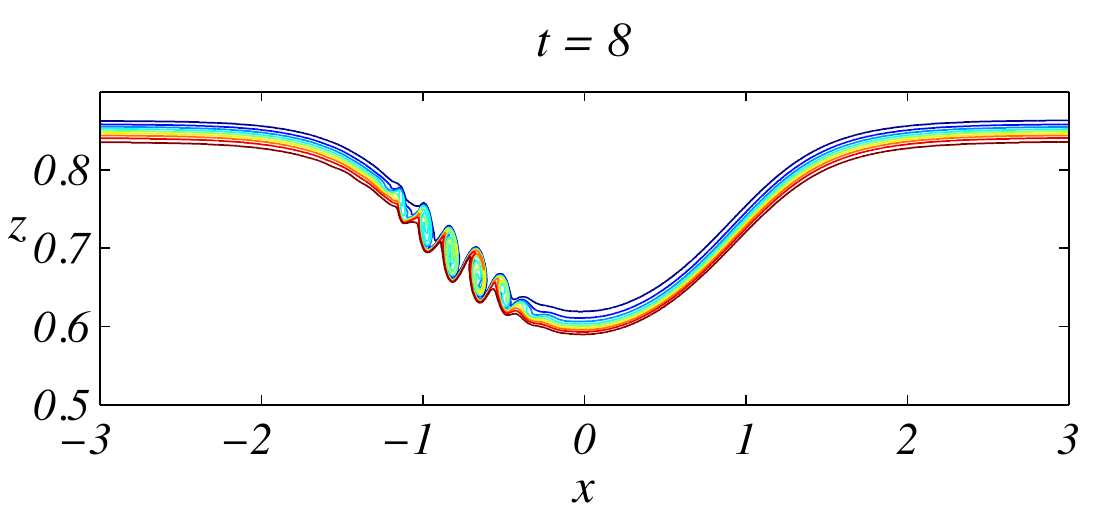}\put(-155,35){(i)}\\ \vspace{3mm}
\includegraphics[width=65mm]{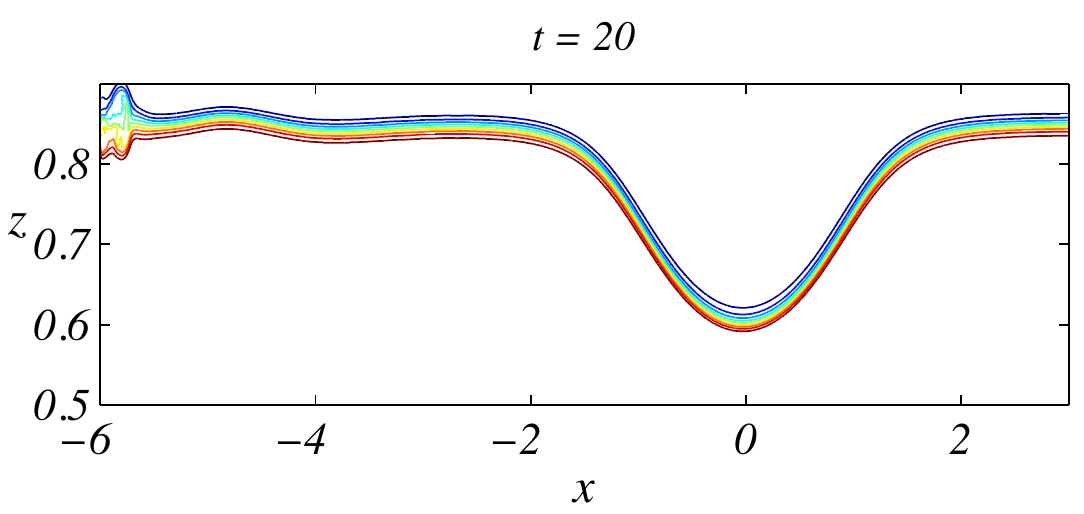}\put(-155,35){(e)}\includegraphics[width=65mm]{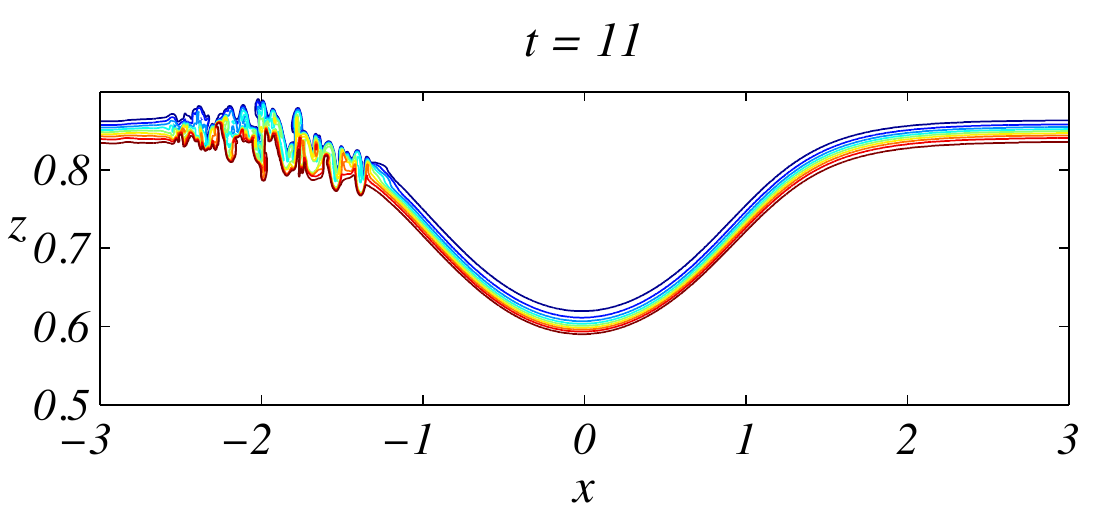}\put(-155,35){(j)}
\caption{(a-e) The nonlinear evolution of the optimal disturbance for the $c=0.4810$ DJL wave. (f-j) Same as (a-e) except the initial disturbance is a Gaussian packet of linear internal waves with $\omega = -14.85$ and $k_+=38.4$ initially centered at $x=3$. In both cases $E_0/E_{ISW}(0) = 10^{-5}$. The panels show contours of the density $S = [0.1:0.1:0.9]$ at the indicated times. Note that only part of the model domain is shown.}
\label{fig_isws020-021}
\end{center}
\end{figure}

As an example of the model fidelity, a run initialized with the $c=0.4810$ DJL wave and no disturbance was integrated for $25$ time units after which the change in domain integrated energy of the solitary wave, $\Delta E_{ISW}(25)/E_{ISW}(0)= 2.3\times10^{-4}$. The loss is nearly constant in time and not a consequence of a rapid initial adjustment of the Boussinesq solitary wave to the non-Boussinesq numerical model. The integration time is relatively short. However, in the calculations presented next, comparable integrations time are sufficient for disturbances to be swept from the domain. 

The evolution of the $c=0.4810$ ISW seeded with the DAL optimal disturbance is shown in Figure \ref{fig_isws020-021}(a-e).  This disturbance is the same as shown in Figure \ref{fig_TOG_structure_c4810}. The panels show contours of the density field, $s(x,z,t)$, at the indicated times. The initial optimal disturbance in this and subsequent calculations has $\Rey=10^5$ since the structure of the optimal disturbances was found to be only weakly dependent on $\Rey$ in this range. The ratio of the initial energy of the perturbation, $E_0$ from (\ref{energy_dal}), to the ISW energy, $E_0/E_{ISW}(0) = 10^{-5}$. In calculating $E_0$, $N_*^2$ is replaced with the local buoyancy frequency field of the ISW, $N^2 = -S_z$. The energy of the perturbation corresponds to $\max |u'|/c = 7.4 \times10^{-2}$ and thus is a relatively large perturbation chosen to produce a large response. However, the maximum perturbation isopycnal displacement $\max(|\eta'|) = 6.1 \times 10^{-5}$, is very small and the disturbance is not apparent in a plot of the density field at $t=0$ (not shown). As with previous nonlinear calculations of this type, the disturbance grows to finite-amplitude K-H billows that breakdown into turbulence as the packet leaves the unstable zone. The turbulence and mixing in the packet evident for $t\ge4$ are certainly not captured correctly in these two-dimensional calculations; however, \citet{BaradF:10} found that three-dimensional effects did not become important until after the disturbances exits the primary ISW. Thus these calculations should give reliable estimates of energy loss from the ISW, but not capture the ultimate fate of the turbulent billows and resulting vertical mixing. Note also that after the disturbance packet leaves the wave, internal waves are radiated behind the ISW ($t=20$) as the ISW adjusts. The total energy loss at $T=25$, after the turbulent patch and the radiated waves have exited the domain, is $\Delta E_{ISW}(25)/E_{ISW}(0)= 1.6\times10^{-2}$. This is two-orders of magnitude larger than the loss associated with the numerical scheme over the same period. 

For comparison, Figure \ref{fig_isws020-021}(f-j) shows the same ISW initiated with a free linear wave packet with frequency $\omega=-14.85$ and $k_+=38.4$ (see Figure \ref{fig_linwave_FWD_c4810}) centered at $x=3$. The disturbance also has $E_0/E_{ISW}(0) = 10^{-5}$, although $\max |u'|/c = 1.5 \times10^{-2}$ and $\max(|\eta'|) = 3.1 \times 10^{-3}$ are different. The larger initial isopycnal displace reflects the equipartition of kinetic and potential energy in the free wave packet. Again, the disturbance grows to produce large billows and turbulence, although they are clearly weaker than in Figure \ref{fig_isws020-021}(a-e). The energy loss $\Delta E_{ISW}(25)/E_{ISW}(0)= 5.0\times10^{-3}$ is only about one third of the loss due to the optimal disturbance. 

The consequences of non-normality are clearly evident in the comparison. The optimal disturbance achieves finite amplitude, i.e. overturning isopycnals, before reaching the mid-point of the wave (panels a and b) while this only occurs after the mid-point of the ISW for the linear wave packet (panel h). This might be attributed to the relatively large size of the initial optimal disturbance. However, the Gaussian wave packet is initially equally energetic. Furthermore, an optimal disturbance with $E_0/E_{ISW}(0) = 10^{-7}$ also produces overturning billows on the leading face of the ISW, while equivalent energy free wave packet does not produce a measurable energy loss.

\begin{figure}
\begin{center}
\includegraphics[width=80mm]{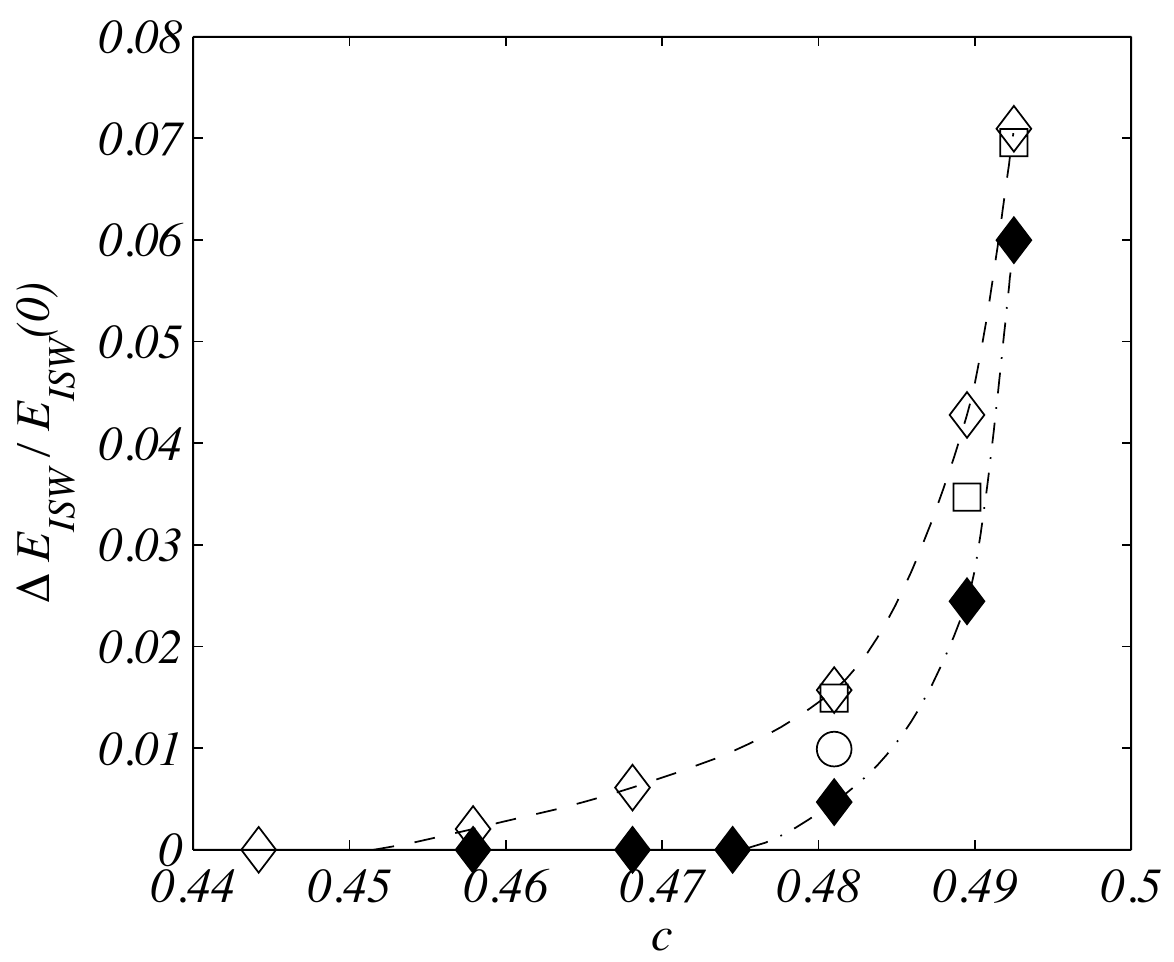}
\caption{Fractional ISW energy loss as a function of $c$ from one disturbance packet. The initial amplitude of perturbation $E_0/E_{ISW}=10^{-7}$ (circles), $10^{-6}$ (squares), and $10^{-5}$ (diamonds). The open symbols are for DAL optimal disturbances and the closed symbols are for Gaussian free linear wave packets. The dashed (dash-dot) line shows $\gamma(c)$ fit to the $E_0/E_{ISW}=10^{-5}$ data for the optimal (free wave) packet.}
\label{fig_dE_vs_c}
\end{center}
\end{figure}

The ISW energy losses from single optimal disturbance packet as a function of $c$ and $E_0/E_{ISW}(0)$ are shown in Figure \ref{fig_dE_vs_c}. Finite energy loss for $E_0/E_{ISW}(0) = 10^{-5}$ is found for $c \gtrsim 0.45$, where $Ri_{min} \approx 0.2$, and $L_{Ri}/\xi \approx 0.43$ and is $\approx7$\% for $c=0.4925$, the largest ISW considered. Although not exhaustive, the results indicate that the energy loss saturates for $E_0/E_{ISW}(0) \approx 10^{-5}$. The corresponding results for the Gaussian free wave packets with $E_0/E_{ISW}(0) \approx 10^{-5}$ are also shown. For these disturbances, finite energy loss occurs for $c\gtrsim0.475$, where  $Ri_{min} \approx 0.11$, and $L_{Ri}/\xi \approx 0.81$, and reaches about $6$\% for $c=0.4925$. The difference between the optimal and free wave disturbances increases as $c$ decreases, reflecting the increased significance of non-normal effects in this range (c.f. Figure \ref{fig_logGmax_vs_c}). From these data maximal loss functions, $\gamma(c)$, are found by fitting to the $E_0/E_{ISW}(0) = 10^{-5}$ data for both the optimal and free wave disturbances. These are shown in Figure \ref{fig_dE_vs_c} by the dashed (dash-dot) line for the optimal (free wave) disturbance. Attempts to distinguish between losses due to K-H instability and the radiated waves proved unreliable. The calculations do show that the wave radiation becomes increasingly significant as $c$ increases.

\subsection{Long-time evolution of ISWs forced by optimal perturbations}

The evolution of base states subjected to continuous, periodic forcing with spatial structure given by the optimal perturbation has been considered by 
\citet{brandt:11} and \citet{sipp:13}. However, a simpler, approximate approach that follows from models of adiabatic decay of KdV solitary waves \citep{Grimshaw:03} is possible. From the previous section, each encounter of a disturbance packet with a wave extracts a small fraction of energy from the ISW given by $\gamma(c)$. Assuming a slow, adiabatic adjustment to each encounter, the ISW energy evolves as
\begin{equation}
\frac{d E_{ISW}}{dx}  \approx -N_p\gamma(c)E_{ISW},
\label{adiab_model}
\end{equation}
where $c(E_{ISW})$ and $\eta_{MAX}(E_{ISW})$ are found from DJL solution family, and $N_p$ is the number of disturbance packets per unit length. The disturbance packets have a length $\approx 0.5$ (see Figure \ref{fig_TOG_structure_c4810}), so that $N_p = 2$ approximates continuous encounters and thus the maximum rate of decay. Cases with $N_p<2$ can be obtained by a simple rescaling of $x$.

\begin{figure}
\begin{center}
\includegraphics[width=80mm]{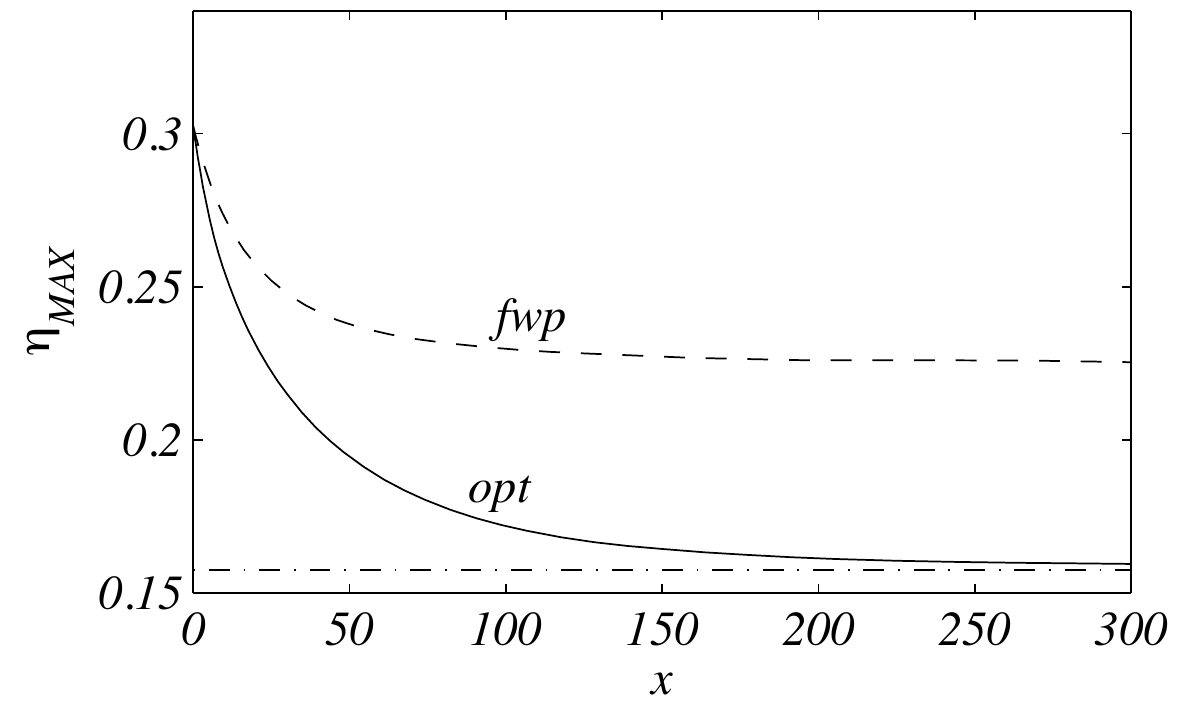}
\caption{Slowly varying estimates of $\eta_{MAX}(x)$ from (\ref{adiab_model}) with $\eta_{MAX}(0)=0.49$ ($c=0.49$), $N_p=2$ and $\gamma(c)$ from the optimal (solid line) or free wave (dashed line) disturbances. The dash-dot line shows the largest wave stable to optimal disturbances at $\eta_{MAX}=0.158$ and $c=0.451$.}
\label{fig_adiabatic_model}
\end{center}
\end{figure}

Figure \ref{fig_adiabatic_model} shows solutions for the maximal decay, $N_p=2$, of an initial wave with $\eta_{MAX}= 0.3025$ ($c=0.49$) subject to both optimal disturbance and free wave packets. For the optimal disturbance the decay $90$\% of the way to the largest stable wave occurs on a spatial scale $x_{diss} \approx 100$ and time scale $t_{diss} \approx 200$. The free wave decays occurs slightly sooner, although the final waves are quite different. As a consequence, optimal disturbances can result in substantially more total loss of energy from the initial solitary wave. In this example optimal disturbances extract $82\%$ of the initial ISW energy compared to the $52\%$ for the free wave packets. Taking $H=100$ m and $\Delta \rho = 2$ kg m$^{-3}$ as representative of coastal settings, the initial wave has a total energy of $1.71\times10^6$ J m$^{-1}$ and the dimensional decay scales become $x_{diss} \approx 10$ km and $t_{diss} \approx 4$ hr. The rate of dissipation per unit width, $cN_p\gamma E_{ISW}$, falls from $\approx 1100$ W m$^{-1}$ at $x=0$, to $\approx 40$ W m$^{-1}$ at $x = x_{diss}/2$. This initial rate is extremely large, although the later rate is comparable to observational estimates of $10-50$ W m$^{-1}$  \citep{Moum:07, Shroyer:10}. 

\section{Discussion}\label{sec:conclusion}

The linear stability of finite-amplitude internal solitary waves in nearly two-layered stratification has been explored using the method of optimal linear transient growth. The approach determines the structure and properties of disturbances that produce the maximum possible gain of perturbation energy over a finite time horizon. The optimal disturbances take the form of localized wave packets initially located in the interfacial region just upstream of the zone of $Ri < 0.25$. They are tilted into the background shear to take advantage of an initial phase of non-normal growth through the Orr mechanism. As they propagate through the primary wave, the packets remain compact and wave-like with well-defined frequency and carrier wavenumber, while experiencing total energy gains of up to $10^{17}$ for the largest ISW considered. 

The growth and properties of the optimal disturbances were compared to a slowly varying, WKB analysis of spatially growing disturbances of Taylor-Goldstein equation. Agreement between the optimal and WKB disturbance properties (real wavelength $k_r$, carrier frequency $\omega$, packet propagation $x(t)$, and energy gain $G$) was quite good, with the exception of the effects of the initial phase of non-normal growth absent in the normal stability analysis underlying the WKB approach. Interestingly, the extra amount of this non-normal growth was nearly constant, regardless of the ISW phase speed and became an increasing fraction of the total growth as the solitary wave speed $c$, decreased. Further comparison with disturbances consisting of packets of linear free waves with carrier frequencies equal to the optimal transient growth disturbance packet further highlighted the role of non-normal effects. However, in this case, the absorption of the perturbation energy by the primary wave occurred in the leading face of the wave, after which perturbation growthrate and properties mirrored the WKB and optimal disturbance results. Together, these three types of disturbances illustrate that the primary instability is due to spatially growing Kelvin-Helmholtz modes, but that non-normal effects in the leading face of the primary wave play a significant role in determining the total energy gain experienced by an upstream disturbance as it propagates through the ISW.

These differences between disturbance type are significant when considering the finite-amplitude evolution. Nonlinear calculations initiated with optimal disturbances resulted in the development of finite-amplitude Kelvin-Helmholtz billows on the upstream face of the ISW, while comparable packets of free wave only resulted in large billows apparent only on the rear face. As a result, significantly more energy was lost from an ISW forced with an optimal disturbance.  These results conflict with earlier conclusion that instability resulting in finite-amplitude billows requires $\ln(G) > 8-10$ (from $\bar{\omega}_i T_W > 5$ or $2 \bar{k}_i L_{Ri} > 4$), $Ri_{min} \lesssim 0.1$ and $L_{Ri}/\xi \gtrsim 0.8$ \citep{TroyK05,Fructus:09,LambF11}. For the current ISW wave family investigated, that occurs for $c \ge 0.4792$, where $L_{Ri}/\xi \ge 0.923$. The nonlinear calculations for ISW energy loss (Figure \ref{fig_dE_vs_c}) agree with the criterion for free wave disturbances. However, optimal disturbances lead to finite amplitude overturning billows and energy loss for $c=0.458$, where  $Ri_{min} = 0.16$, $L_{Ri}/\xi = 0.60$ and $\ln(G_{opt}) = 5.47$, well below the semi-empirical criteria. Of course, this depends on some measure of the amplitude of the disturbance and more significantly on the presence of an optimal disturbance. However, the ocean thermocline is typically full of energetic motions ranging from random turbulence to free internal waves. It seems reasonable to expect that some of these motions will project onto the optimal structure and lead to large energy growth.

Results for the viscously adjusted DJL waves show that transient growth is only weakly affected at $\Rey = 10^5$ representative of laboratory experiments. In the recent experiments by \citet{Carr:17}, their $141010$ experiment had $H = 75$ cm and a nominal upper layer depth of $10.5$ cm, giving $z_0 = 0.86$, close to our value of $0.85$. A value of $\lambda \approx 80$ for the 
hyperbolic tangent 
density profile (\ref{Sbar}) gives a reasonable match to the density profile and the error function fit shown in their Figure $6$. From their Figure $5$ we estimate a wave amplitude of $23.0$ cm (the average of the two curves shown). This gives $\eta_{MAX}=0.31$, which corresponds to an inviscid DJL wave with $c=0.49$. The average wavelength of the interfacial disturbances shown in their Figure $3$ is $l \approx 15.4$ cm. In our scaling this gives wavenumber $k=2\pi H l^{-1}= 30.6$. The optimal and WKB analysis for the $c=0.4895$ ISW give, respectively, $k = 31.6$ and $31.9$. The agreement between the laboratory result and the prediction from the theory is quite good. Carr {\it et al.}'s experiments show that the billows reach finite amplitude (overturning) at the wave crest, which might indicate a free wave, or K-H, disturbance, although without information on upstream disturbance amplitude, it is not possible to distinguish the origin of the excitation.

\begin{figure}
\begin{center}
\includegraphics[scale=0.6]{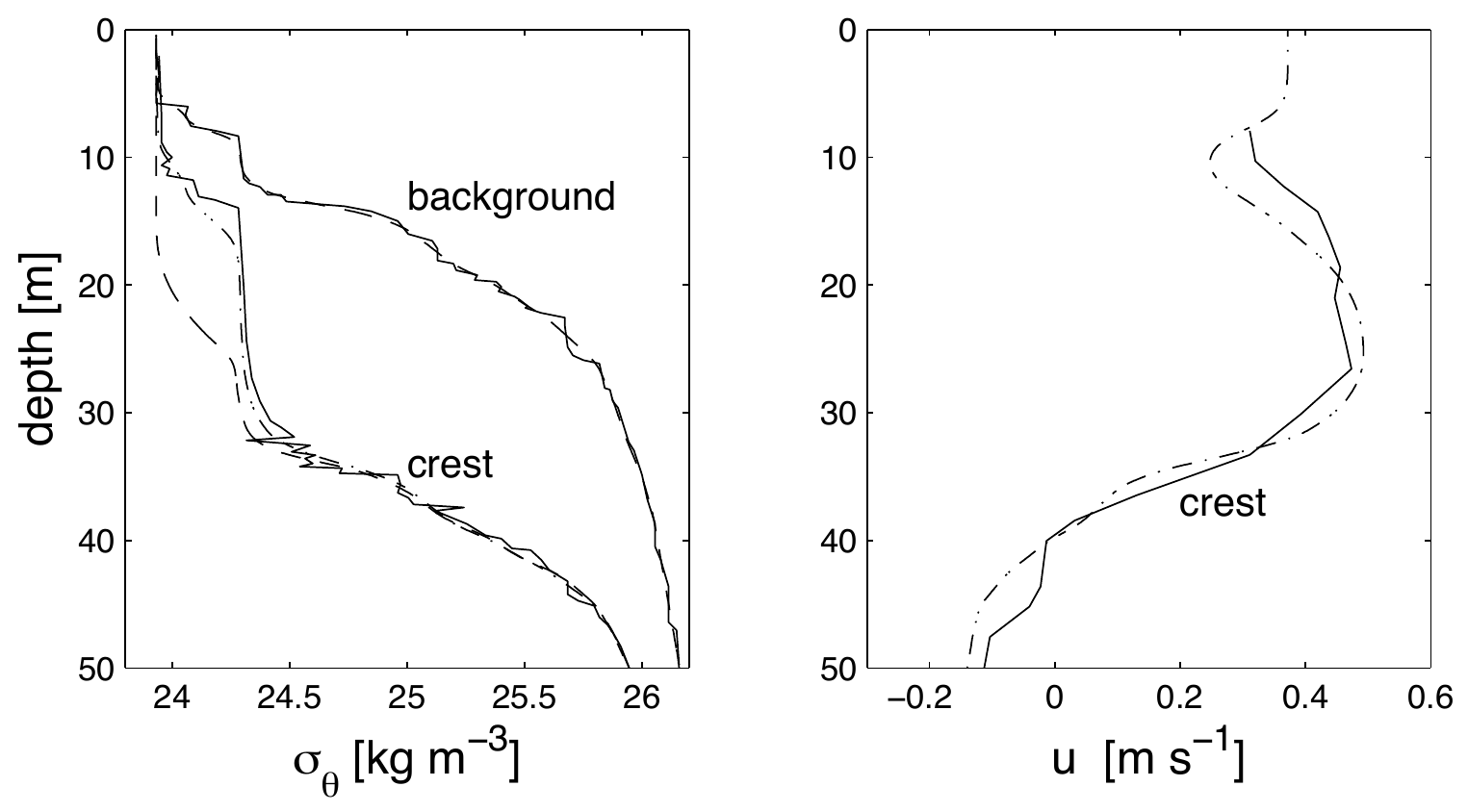}
\put(-175,125){(a)}
\put(-85,125){(b)}
\caption{a) Observed background and ISW crest density profiles from \cite{Moum:03} (solid lines). DJL solitary wave calculations for a still (dashed) and sheared (dash-dot) ambient background flow. b) Observed ISW crest horizontal velocity profile (solid) and the DJL solution with the sheared ambient background flow (dash-dot). See the text for details.}
\label{fig_Moum_etal_2003}
\end{center}
\end{figure}

\cite{Moum:03} tracked trains of ISWs over the Oregon shelf and reported field observations of instabilities and turbulence within the interfacial regions. In particular their Figures $5$ and $6$ show acoustic returns from two realizations of the same wave separated by about 1 hour. In both figures a thin region of high acoustic return indicative of active finite-amplitude overturning and turbulence is present. This signal begins ahead of the wave crest, increases in intensity through the wave and extends behind the wave. The appearance of finite-amplitude disturbances upstream of the crest is consistent with excitation by optimal disturbances (or noise that projects onto them) rather than K-H instability excited by free linear waves or simple harmonic forcing.  

To explore this interpretation further, the upstream density profile in their Figure 8 was used to compute DJL solitary waves in water with total depth $H=98$ m. Figure \ref{fig_Moum_etal_2003}a shows the observed density profiles (digitized from their Figure 8) upstream and at the wave crest. The figure also shows the smoothed version of the upstream profile (dashed) used to compute the DJL families. DJL solitary waves with maximum dimensional isopycnal displacement $\eta_{MAX}=20.4$ m, consistent with the value of $\approx 20$ m in Moum {\it et al}'s Figure 8, were obtained. \cite{Moum:03} mention the presence of a background flow, $u_0(z)$, and use it in their estimates of wave stability, but do not show the profile. Thus the first DJL wave was calculated for $u_0(z)=0$. The density profile at the wave crest is shown by the dashed line in Figure \ref{fig_Moum_etal_2003}a. The wave speed $c = 0.62$ m s$^{-1}$ is in good agreement with the observed value of $0.6$ m s$^{-1}$. The agreement between the crest density profiles below $30$ m depth is very good; however, the profiles disagree between depths of $12$ and $30$ m. This could be a consequence of the background flow so the second family was computed with the flow given by a simple hyperbolic tangent profile  
$$
u_0(z) = \frac{\Delta U}{2} \left( 1 + \tanh \left[ \frac{z-z_0}{d} \right] \right).
$$
Note that in our notation the $z$ origin is at the bottom. (See \cite{StastnaL:02} for the DJL theory with a sheared ambient flow.) The background density profile suggests a surface intensified flow, thus we took $z_0 = 90.2$ m and $d = 1.63$ m, which mirror the sharp ambient density jump centered at a depth of $\approx -8$ m. A retrograde upper layer velocity $\Delta U /2 = -0.144$ m s$^{-1}$ ($= -0.1$ in scaled variables) is suggested by the observed crest density profile, although the magnitude was chosen arbitrarily, but turned out to be a fortuitous guess. The resulting wave crest density profile is shown by the dash-dot line in Figure \ref{fig_Moum_etal_2003}a. The agreement with the observations is now quite good. This is further demonstrated in Figure \ref{fig_Moum_etal_2003}b where the observed wave crest horizontal velocity profile digitized from Figure 18b in \cite{Moum:03} is shown with the profile from the DJL solution with the ambient shear. The DJL solution has been shifted by the addition of a barotropic component of $0.05$ m s$^{-1}$. With this additional mean flow (which does not affect the density structure) the wave speed $c = 0.61$ m s$^{-1}$ is close to the observed estimate. Additional adjustment of the ambient flow could improve the comparison, but does not seem warranted. 

Interestingly, for the case without shear, the minimum Richardson number $Ri_{min} = 0.258$ (at a  depth of $33.7$ m), suggesting that the wave would be stable. However, with the ambient flow included, there are now two zones of low $Ri$, both with $Ri_{min} = 0.19$. They are centered at depths of $13$ and $33.3$ m, coincident with regions of overturns in the observed crest density profile. Of particular interest is that $Ri_{min}$ is not deeply unstable, and in the case of the deeper zone, the length scale ratio $L_{Ri}/\xi = 0.39$ is small. These values suggest instability excitation by optimal disturbances. 
 
Field echo sounder observations \citep{Moum:03,Lien:14} suggest that in the case of trains of ISWs, the leading wave (usually the largest amplitude wave in the train) amplifies small amplitude perturbations to large amplitudes. These perturbations remain energetic and enter the smaller trailing ISWs where they result in observable turbulence. This scenario is especially evident in Figure 3 of \citet{Moum:03}. Transient growth, possibly by noise that projects onto optimal disturbances, appears to be the origin of the transition to turbulence and mixing induced by large amplitude ISWs.

Transient growth instabilities, with the potential for excitation of optimal disturbances, appear to dominate ISWs with sharp interfaces. However, other types of waves such as trapped-core waves (the type-II waves of \cite{ZhangA:15}) and mode-two waves \citep{Shroyer2:10} may also be susceptible to additional instability mechanisms associated closely with the recirculating cores \citep{HelfrichW:10,Carr:12}.

\subsection*{Acknowledgements}
PYP and BLW acknowledge the support by the National Science Foundation Grant Number OCE-1155558. KRH acknowledges support from Independent Research and Development and Investment in Science Program awards from the Woods Hole Oceanographic Institution.

\bibliographystyle{jfm}

\bibliography{bib}

\end{document}